\documentclass[aps,prx,twocolumn,superscriptaddress,showpacs,floatfix]{revtex4-1}
\usepackage{amsmath,amssymb,amsfonts,float,graphics,epsfig,epstopdf,color,verbatim,tabularx,bm,multirow}
\usepackage{hyperref}
\hypersetup{
colorlinks = true,
linkcolor = [rgb]{0.70,0.13,0.13},
citecolor = [rgb]{0.13,0.55,0.13},
urlcolor = [rgb]{0.25,0.41,0.88}}

\DeclareSymbolFont{sfletters}{OML}{cmbrm}{m}{it}
\DeclareMathSymbol{\sfeps}{\mathord}{sfletters}{"22}

\graphicspath{{Figures/}}

\begin{document}

\title{Metal to Orthogonal Metal Transition}

\author{Chuang Chen}
\affiliation{Beijing National Laboratory for Condensed Matter Physics and Institute of Physics, Chinese Academy of Sciences, Beijing 100190, China}
\author{Xiao Yan Xu}
\affiliation{Department of Physics, Hong Kong University of Science and Technology, Clear Water Bay, Hong Kong SAR, China}
\affiliation{Department of Physics, University of California at San Diego, La Jolla, California 92093, USA}
\author{Yang Qi}
\email{qiyang@fudan.edu.cn}
\affiliation{Center for Field Theory
and Particle Physics, Department of Physics, Fudan University,
Shanghai 200433, China}
\affiliation{State Key Laboratory of Surface Physics, Fudan University, Shanghai 200433, China}
\affiliation{Collaborative Innovation Center of Advanced
Microstructures, Nanjing 210093, China}
\author{Zi Yang Meng}
\email{zymeng@hku.hk}
\affiliation{Department of Physics and HKU-UCAS Joint Institute of Theoretical and Computational Physics, The University of Hong Kong, Pokfulam Road, Hong Kong SAR, China}
\affiliation{Beijing National Laboratory for Condensed Matter Physics and Institute of Physics, Chinese Academy of Sciences, Beijing 100190, China}
\affiliation{Songshan Lake Materials Laboratory, Dongguan, Guangdong 523808, China}
\date{March 27, 2020}

\begin{abstract}
Orthogonal metal~\cite{Nandkishore2012,Kaul2012,Reuegg2010} is a new quantum metallic state that conducts electricity but acquires no Fermi surface (FS) or quasiparticles, and hence orthogonal to the established paradigm of Landau's Fermi-liquid (FL). Such state might hold the key of understanding the perplexing experimental observations of quantum metals that are beyond FL -- dubbed non-Fermi-liquid (nFL) -- ranging from the Cu- and Fe-based oxides~\cite{Loehneysen2007,Keimer2015,Badoux2016,YHGu2017}, heavy fermion compounds~\cite{Stewart2001,Custers2003,QMSi2010,Steppke2013} to the recently discovered twisted graphene heterostructures~\cite{cao2018correlated,cao2018unconventional,YuanCao2019,ChengShen2019}. However, to fully understand such exotic state of matter, at least theoretically, one would like to construct a lattice model and solve it with unbiased quantum many-body machinery. Here, we achieve this goal by designing a 2D lattice model comprised of fermionic and bosonic matter fields coupled with dynamic $\mathbb{Z}_2$ gauge fields, and obtain its exact properties with sign-free quantum Monte Carlo simulations. We find as the bosonic matter fields become disordered, with the help of deconfinement of the $\mathbb{Z}_2$ gauge fields, the system reacts with changing its nature from the conventional normal metal with a FS to an orthogonal metal of nFL without FS and quasiparticles and yet still responds to magnetic probe like a FL. Such a quantum phase transition from a normal metal to an orthogonal metal, with its electronic and magnetic spectral properties revealed, is calling for the establishment of new paradigm of quantum metals and their transition with conventional ones.
\end{abstract}

\maketitle

\section{Introduction}

As one cornerstone in condensed matter physics, Landau's Fermi liquid (FL) theory teaches us that at zero temperature, a Fermi liquid has a Fermi surface (FS) marked by the momenta of gapless quasiparticle excitations, similar to its noninteracting counterpart. When the electron number is held fixed, the volume inside the FS is invariant upon interaction. This is the statement given by Luttinger at 1960~\cite{Luttinger1960}, and by now the perturbative argument has become the {\it Luttinger's theorem} and later Oshikawa modernized the argument from a topological perspective~\cite{Oshikawa2000,Paramekanti2004}. Under these guidelines, the volume inside the FS is conserved even in an interacting FL, and the reduction of FS must come from the breaking of translational symmetry which enlarges the elementary unit cell of the problem at hand.

\begin{figure*}[htp!]
\begin{center}
\includegraphics[width=0.8\textwidth]{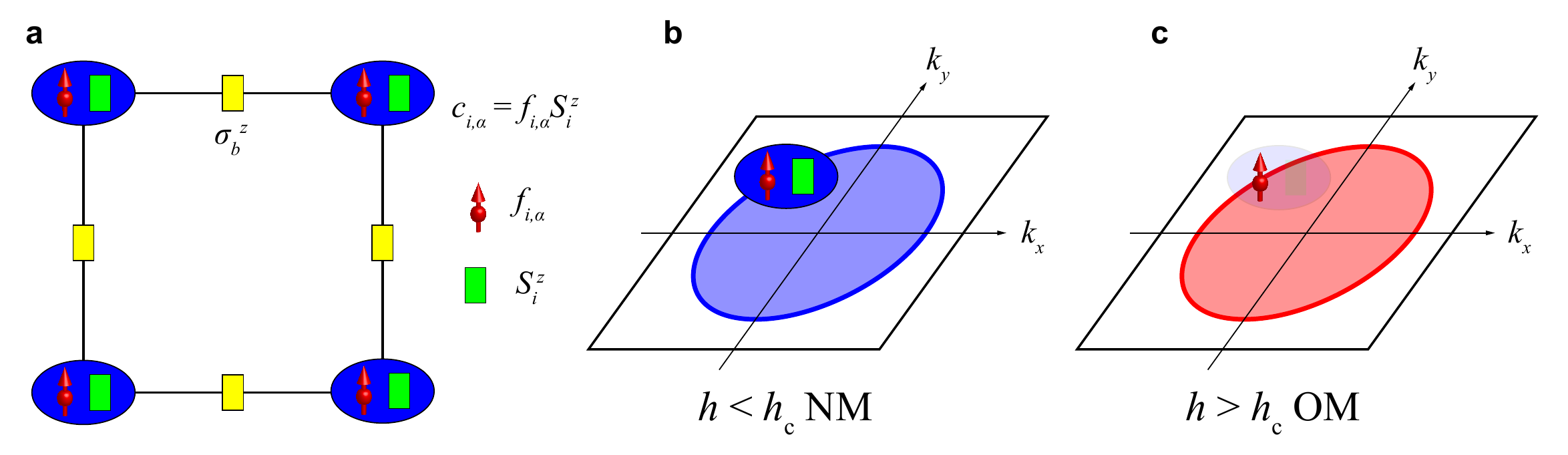}
\caption{Metal and its awkward cousin. (a) The lattice model in Eq.~\eqref{eq:H}, on a square lattice, there are composite fermions $c_{i,\alpha}=f_{i,\alpha}S^{z}_{i}$ on each site $i$, comprised of orthogonal fermion field $f_{i,\alpha}$ and Ising matter field $S^{z}_{i}$. The $\mathbb{Z}_{2}$ gauge field $\sigma^{z}_{b}$ lives on the bond $b$. The blue ellipse stands for the situation in which the composite fermion is a well-defined quasiparticle. (b) The FS of the system inside the NM phase $(h<h_c)$. The blue circle encloses the area of FS corresponding to the density of $c$ fermions. This is consistent with the Luttinger's theorem. (c) The hidden FS of the system inside the OM phase $(h>h_c)$. The red circle encloses the same areas as the blue one in (b) but since the $c$ fermions lose coherence inside OM, the red FS cannot be detected from single-particle spectral probes. The quantum phase transition from (b) to (c) signifies the breakdown of the Luttinger's theorem without symmetry breaking.}
\label{fig:fig1}
\end{center}
\end{figure*}

Interestingly, as it is often happened in physics, experimental discoveries could be well ahead of theoretical understandings. By now, there are ample examples of correlated electron systems that share the deviant behavior that strongly violates the relation between the volume of quasiparticle FS and the electron filling. These systems are in general dubbed as non-Fermi liquid (nFL) -- ranging from the Cu-, Fe-, Cr- and Mn-based superconductors~\cite{Loehneysen2007,Keimer2015,YHGu2017,Wu2014,Cheng2017,JGCheng2018}, heavy fermion compounds~\cite{Stewart2001,Custers2003,QMSi2010,Steppke2013}, to the recently discovered twisted graphene heterostructures~\cite{cao2018correlated,cao2018unconventional,YuanCao2019,ChengShen2019}. Although it is generally accepted that their behavior is a result of electron-electron interaction and perhaps disorder, but despite many proposals over the decades, such as the fractionalized FL~\cite{Paramekanti2004}, FL$^{*}$ phase~\cite{Senthil2003,Punk2015,Feldmeier2018} and SYK type of nFL~\cite{MaldacenaStanford2016}, which are shown to exist by recent quantum Monte Carlo simulation~\cite{Hofmann2018, GPPan2020}, there exist no universally accepted theory that could describe their behavior. While explaining experimental observations is the ultimate goal of any theory, as a first step, simple lattice models that have metallic ground states but do not fall into the FL theory and manifest no quasiparticle FS are highly desirable. Even such level of model and its unbaised solution, in correlated electron systems higher than 1D, does not exist.

Whether the realization of a nFL in a correlated electronic model with either reconstruction or complete destruction of FS without symmetry breaking can be realized in a concrete manner, is the question we address in this work. Here we show, that such a nFL model with no quasiparticle FS can be constructed in a correlate electron system and solved with unbiased large-scale quantum Monte Carlo simulations. The destruction of the entire FS can indeed happen without any symmetry-breaking, and a continuous quantum phase transition from normal metal (NM) of FL to an orthogonal metal (OM) of nFL manifests. The OM phase thence discovered, is a new state of quantum metal beyond the establishing paradigms in condensed matter physics.
In particular, it differs from the FL$^\ast$ phase as the former has a hidden FS formed by fractionalized fermionic excitations carrying electric charges and spins, while the latter has a conventional FS made of conventional quasiparticles, coexisting with a gapped topological order containing only gapped fractionalized excitations.

\begin{figure*}[htp!]
\begin{center}
\includegraphics[width=0.8\textwidth]{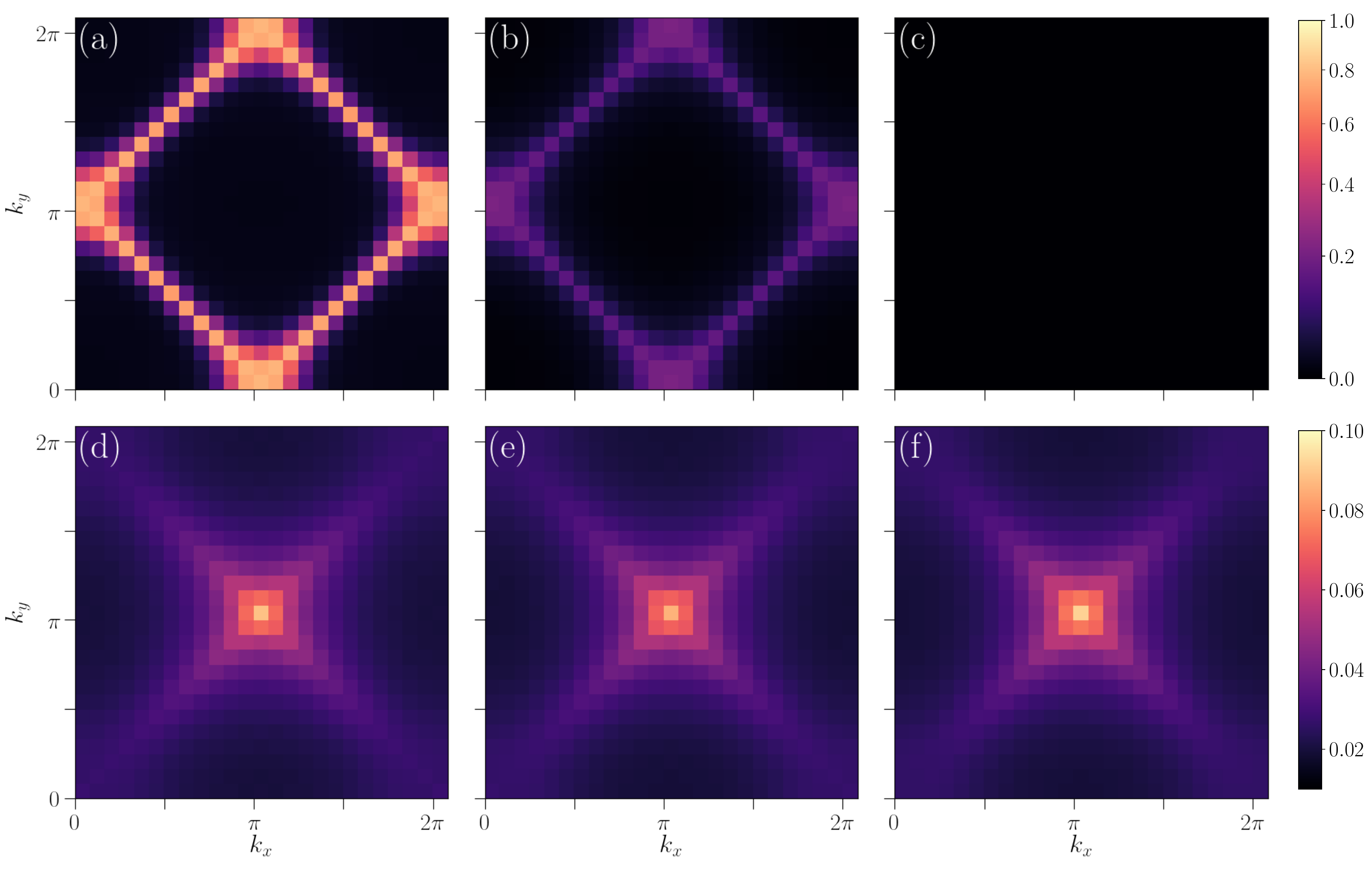}
\caption{Fermionic and bosonic responses. (a), (b) and (c): the spectra at the FS $A(\mathbf{k},\omega=0)\propto G(\mathbf{k},\beta/2)$ of composite fermion $c$ for $L=24$, $T=0.1$, $g=0.5$ systems. For $h=0.4$ ((a), $h<h_c$, NM), $h=2$ ((b), $h\sim h_c$, QCP) and $h=4$ ((c), $h>h_c$, OM). (d), (e) and (f): spin susceptibility $\chi(\mathbf{q},\omega=0)$ for the same parameter sets. It is clear that in the NM phase (a), the dimond shape FS gives rise to the magnetic instability at $\mathbf{q}=(\pi,\pi)$ in (d), but as the NM evolves into OM, the FS vanishes as shown in  (b) and (c), its magnetic response does not change in any obvious way, (e) and (f), that, there still exists the instability at $\mathbf{q}=(\pi,\pi)$ despite of the fact that there is no FS to be nested. This is the special properties of the OM that it responds like a metal (magnetically and electronically) but there is a gap in its $A(\mathbf{k},\omega=0)$.}
\label{fig:fig2}
\end{center}
\end{figure*}

\section{Model, Message and Method}
Our model, inspired by the proposals of orthogonal fermion construction~\cite{Nandkishore2012,Reuegg2010,Gazit2019}, has the following Hamiltonian on a 2D square lattice, $H=H_{f}+H_{z}+H_{g}$, where
\begin{eqnarray}
H_f &=& -t\sum_{\langle i,j \rangle} (f^{\dagger}_{i,\alpha} \sigma^{z}_{b_{\langle i,j \rangle}}f_{j,\alpha} + h.c.) -\mu\sum_{i}f^{\dagger}_{i,\alpha}f_{i,\alpha}, \nonumber\\
H_{z} &=& -J \sum_{\langle i,j \rangle} S^{z}_{i} \sigma^{z}_{b_{\langle i,j \rangle}} S^{z}_{j} - h \sum_{i} S^{x}_{i}, \nonumber\\
H_{g} &=& -K \sum_{\square}\prod_{b\in\square} \sigma^{z}_{b} - g\sum_{b} \sigma^{x}_{b}
\label{eq:H}.
\end{eqnarray}
The model is depicted in Fig.~\ref{fig:fig1} (a) with the parameters simplified in the following manner: for the $f$ orthogonal fermion part $H_{f}$, we set its nearest neighbor hopping amplitude $t=1$, and the chemical potential $\mu=0$ to fix the half-filling of the $f$ fermions (in the Appendix.II, we also show results away from half-filling); for the Ising matter field part $H_{z}$, we set nearest neighbor ferromagnetic interaction $J=1$ and use the transverse field $h$ as the control parameter for the quantum fluctuations; for the $\mathbb{Z}_2$ gauge field part $H_{g}$, we set $K=1$ such that zero flux per plaquette $\square$ is favored, and $g=0.5$ is small enough to not break the $\mathbb{Z}_2$ topological order in $H_{g}$, yet still large enough to provide sufficient gauge fluctuations.

The physical -- gauge neutral -- fermionic degree of freedom in our model, is the composite ($c$) fermion made out of the orthogonal fermion $f$ and Ising matter field $S^{z}$, in that, $c^{\dagger}_{i,\alpha} (c_{i,\alpha}) = f^{\dagger}_{i,\alpha} S^{z}_{i} (f_{i,\alpha}S^{z}_{i})$, denoted as the blue ellipse in Fig.~\ref{fig:fig1} (a) and (b). It is the FS structure of the $c$-fermions that we will pay most of our attention to in this paper, denoted as the blue circle in Fig.~\ref{fig:fig1} (b). And our main finding is that when the $f$ fermions form a metallic state in the presence of $\mathbb{Z}_2$ topological order of the $\sigma$ gauge field and the disordered phase of the $S^{z}$ matter field, the FS of the $c$ composite fermions vanishes with their quasiparticle fraction reduced to zero.
This appears to violate the Luttinger's theorem in having a symmetric metallic state without FS, as illustrated schematically in Fig.~\ref{fig:fig1} (c):
A generalized Luttinger's Theorem is only recovered if a hidden FS of the fractionalized $f$-fermion is also accounted for~\cite{Liza2011,Gazit2019}.
As we tunes the $\mathbb{Z}_2$ gauge field towards confinement, by means of controlling the quantum fluctuations in the Ising matter fields, the entire system goes through a continuous transition after which the FS of the composite fermions is recovered. This is illustrated in Fig.~\ref{fig:fig1} (b).

The phase with $\mathbb{Z}_2$ deconfinement as well as the vanishing of quasiparticle of composite fermions and their FS is a new state of quantum metal~\cite{Senthil2002,Nandkishore2012,Kaul2012,Reuegg2010} (following the literatures, we denote it as orthogonal metal, or OM in short), and the continuous quantum phase transition from it to the normal metal, or NM in short, of composite fermion carrying the flavor of Higgs transition of Ising gauge field~\cite{Kogut1979,Fradkin2013}. In the NM phase, a gauge-neutral string-order in the Ising matter field is developed and the coherent fermionic quasiparticles reappear. These results provide concrete material for future field theoretical analysis.

To be able to solve the Hamiltonian in Eq.~\eqref{eq:H} in an unbiased manner, we employ determinantal quantum Monte Carlo (QMC) simulations. The generic method description of simulating fermionic degrees of freedom coupled to critical bosonic fields can be found in the review~\cite{XiaoYanXu2019} and the simulation performed here is close to the ones in Ref.~\cite{XiaoYanXu2018,YZLiu2020}, with the complexity that now the two sets of fields, Ising matter field $S^{z}_{i}$ and $\mathbb{Z}_{2}$ gauge field $\sigma^{z}_{b}$ have to be updated sequentially, see the Method section for details. The physical observables calculated are all gauge-neutral such as the dynamic Green's function of the composite fermions $G(\mathbf{k},\tau)=\frac{1}{N}\sum_{i,j,\alpha} e^{i\mathbf{k}\cdot\mathbf{r}_{ij}}\langle c^{\dagger}_{i,\alpha}(\tau) c_{j,\alpha} (0) \rangle$ with $N=L^{2}$ and $\tau\in[0,\beta]$. As will be clear in the results section, $G(\mathbf{k},\tau=\beta/2)$ is used to approximate the composite fermion spectral function at the Fermi level $A(\mathbf{k},\omega=0)$ and to extract the quasiparticle fraction $Z_{\mathbf{k}_F}$. Other physical observables are given in the Method section.

We also notice the similarity of our OM phase with that discovered in recent works~\cite{Hohenadler2018,Hohenadler2019,Gazit2019}. The Higgs transition between NM and OM in this work is replaced with a finite temperature crossover in Refs.~\cite{Hohenadler2018,Hohenadler2019}, as in the latter, the system behave as a conventional square lattice Hubbard model at low temperature. Also, in Ref.~\cite{Gazit2019}, via QMC study of an extended model, quantum phase transitions between metals without symmetry breaking is discovered and one of the metallic phase is an orthogonal semi-metal.

\section{Symmetry}
\label{sec:symmetry}
\begin{figure}[htp!]
\begin{center}
\includegraphics[width=\columnwidth]{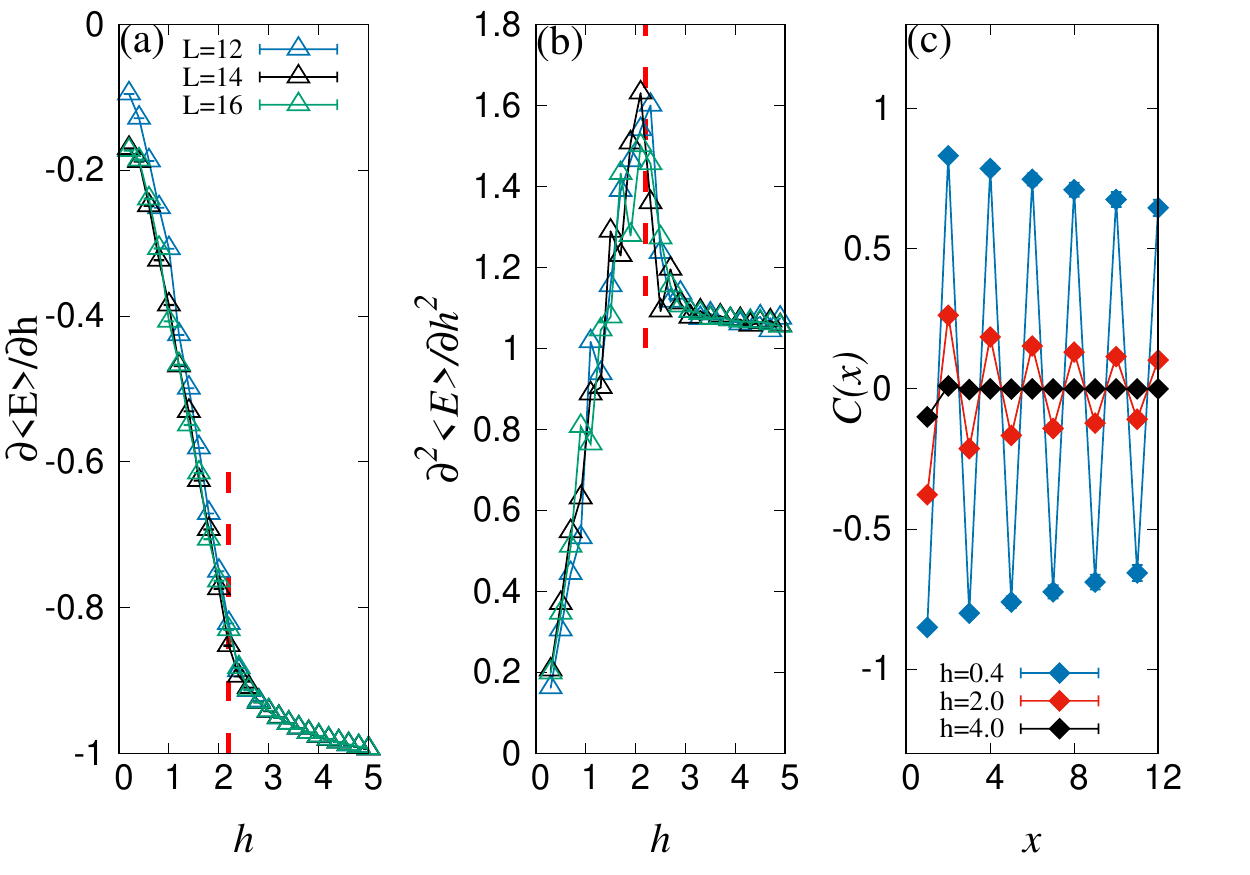}
\caption{QCP and string operator. (a) The internal energy derivative as a function of $h$, it is clear that the energy derivative is a continuous function with change of slope at $h_c =2.2(2)$, denoted by the red dashed line. The results are for $L=10, 12$ and $16$ systems. (b) The second derivative of the internal energy, a peak is seen at $h_c=2.2(2)$, denoted by the red dashed line, consistent with the position in (a). (c) The correlation of string operator $C(\mathbf{r})$ at different $h$ as a function of distance along the $\hat{x}$ lattice direction. Inside the NM phase (blue curve with $h=0.4$), the string operator develops long-range order, and such order is gradually reduced as $h$ approaches $h_c$ (red curve with $h=2$), and inside the OM phase (black curve with $h=4$), the correlation is exponentially small even in a finite size system. These results are obtained from $L=24$, $\beta=10$ system.}
\label{fig:fig3}
\end{center}
\end{figure}

Before presenting the numerical results, we would first like to analyze the symmetry properties acquired by model in Eq.~\eqref{eq:H}.

First, there is a $\mathbb Z_2$ gauge symmetry. The Hamiltonian is invariant under the following gauge transformation $f_{i,\alpha}\rightarrow\eta_if_{i,\alpha}$ and $S_i^z\rightarrow\eta_iS_i^z$, where $\eta_i=\pm1$ is a site-dependent $\mathbb Z_2$ factor.
Correspondingly, we can define local operators $Q_i = (-)^{n_{i,f}}S^{x}_{i}\prod_{b\in +_i}\sigma^{x}_{b}$ with $n_{i,f}=\sum_{\alpha}f^{\dagger}_{i,\alpha}f_{i,\alpha}$ and $+_{i}$ stands for the four bonds oriented from site $i$, which performs the gauge transformation on site $i$.
It can be shown that $[Q_i,H]=0$ for all the sites, so the eigenvalues of $Q_{i}=\pm 1$ are conserved quantities,
and they span an infinite set of local $\mathbb{Z}_2$ gauge invariants.
In particular, we consider the subspace of states satisfying the constraints $Q_i=1$.
They form the Hilbert space of a $\mathbb Z_2$ gauge theory, with even $\mathbb Z_2$ gauge structure~\cite{Paramekanti2004} and the constraints $Q_i=1$ play the role of the Gauss law.
One important consequence of the gauge symmetry is that only gauge-neutral operators can have nonvanishing expectation values.
For example, $\langle c_{i\alpha}^\dagger c_{j\alpha}\rangle$ may be nonzero, but $\langle f_{i\alpha}^\dagger f_{j\alpha}\rangle$ will always be zero, as we shall see later in the results of our simulation.

Second, the model Hamiltonian has a global $\mathbb Z_2$ symmetry $S_i^z\rightarrow-S_i^z$.
In our simulation, the quantum fluctuations in the Ising matter field, controlled by the transverse field $h$ will drive the OM to NM transition which breaks this symmetry in the latter phase. The OM-to-NM transition should be regarded as a Higgs transition related with that in the Ising-Higgs gauge theory~\cite{Fradkin2013}, because when combined with the fermion-parity symmetry $f_{i\alpha}\rightarrow-f_{i\alpha}$, the $\mathbb Z_2$ symmetry operation is realized as a $\mathbb Z_2$ gauge transformation with $\eta_i=-1$ on all sites.
Since the fermion-parity symmetry can never be broken, the breaking of the $\mathbb Z_2$ symmetry is equivalent to the breaking of the combined symmetry, which is a gauge symmetry, results in the Higgs transition.
Being a Higgs transition has two consequences: First, the transition will eliminate the $\mathbb Z_2$ gauge field from the low-energy effective theory in the NM phase. As we shall see later, this means that the transition will terminate the $\mathbb{Z}_2$ topological order in the OM phase, and realizes a traditional NM of Fermi-liquid. Second, it implies that there is no local bosonic order parameter for the symmetry breaking and the probe of such order will rely on the string operator constructed out of the Ising matter field and $\mathbb{Z}_2$ gauge field, even though the topological order in the gauge field is absent.
In other words, after the Higgs transition and inside the NM phase, the $\mathbb Z_2$ symmetry, as a part of the gauge symmetry, cannot be truly broken (because a symmetry-breaking phase is defined by the long-range order of a local bosonic order parameter).

Third, the model has a global U(1) charge-conservation symmetry $f_{i,\alpha}\rightarrow f_{i,\alpha}e^{i\theta}$.
Both $f$ and $c$ fermions carry unit U(1) charge.
The presence of this symmetry allows us to define the filling of the $c$-fermions, and results in the Luttinger's counting in the NM phase.
We shall see that the theorem is violated in the OM phase, despite of the fact that this U(1) symmetry is not broken, in fact none of these three symmetries are broken inside the OM phase and this manifests the non-trivial properties of the OM phase discovered in this paper.

Last, there is a particle-hole symmetry $f_{i,\alpha}\rightarrow\pm f_{i,\alpha}^\dagger$ on the two sublattices, respectively.
The particle-hole symmetry pins the density of the physical fermion $c_i$ at half-filling, when no chemical-potential term is added (in the Appendix II, we show results away from half-filling).

\section{Quantum phase transition between NM and OM}
With the above analyses in mind, we are now ready to discuss the NM-to-OM quantum critical phase transition revealed by our QMC results.

\begin{figure}[htp!]
	\begin{center}
		\includegraphics[width=\columnwidth]{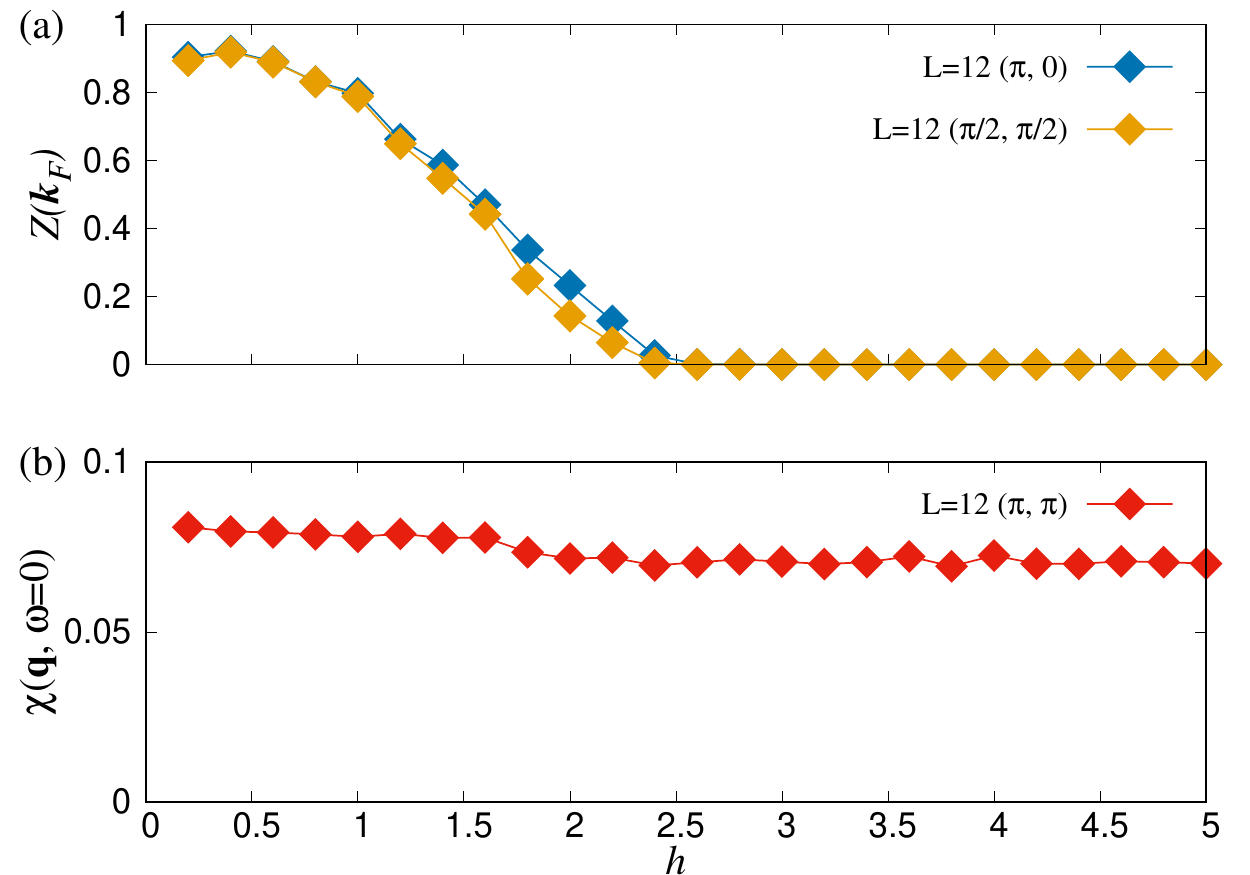}
		\includegraphics[width=.977\columnwidth]{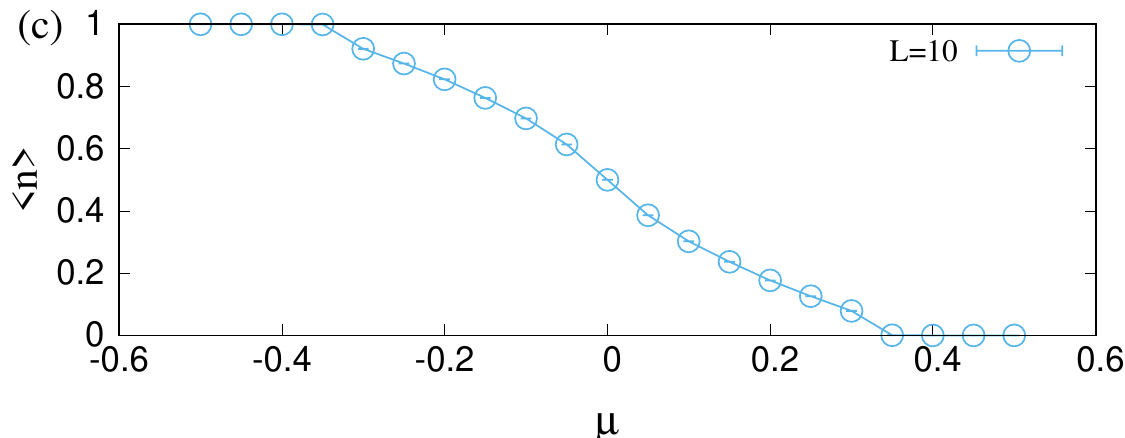}
		\caption{Single particle and magnetic residual. (a) $Z(\mathbf{k}_F)\sim G(\mathbf{k},\beta/2)$ for $L=12$, $\beta=10$ systems as a function of $h$ at two difference momenta $(\pi,0)$ and $(\pi/2,\pi/2)$. $Z(\mathbf{k}_F)$ starts from 1 deep inside the NM phase and gradually reduces to zero at the quantum phase transition between NM and OM. Inside the OM phase $Z(\mathbf{k}_{F})=0$. (b) the magnetic susceptibility $\chi(\mathbf{q},\omega=0)$ with $\mathbf{q}=(\pi,\pi)$. The response is a constant in both phases and at the critical point, meaning that the OM phase although has no single particle residual, is a metal in disguise in its magnetic response. (c) Average charge density as a function of chemical potential: $\langle n \rangle (\mu)$ for $h=4$ inside the OM phase. The charge density $n$ varies continuously with respect to $\mu$, indicating that there is no charge-insulating region.}
		\label{fig:fig4}
	\end{center}
\end{figure}

The most straightforward way to observe this transition is to measure the FS of the composite fermion. The upper panel of Fig.~\ref{fig:fig2} demonstrates the evolution of the FS as one moves along the axis of $h$. The system size is $L=24$ and the inverse temperature $\beta=10$. What is plotted here is the dynamic Green's function of $c$ fermions $G(\mathbf{k},\tau)$ over the Brillioun zone and it can be used to approximate the spectra as $ A(\mathbf{k},\omega=0) \approx \beta G(\mathbf{k},\beta/2)$ in the limit $\beta \to \infty$~\cite{XiaoYanXu2017,ZHLiu2018,ZHLiu2019,Hohenadler2018}.
It is clear that for small $h$ [see Fig.~\ref{fig:fig2} (a)], the FS is identical to that of an non-interacting one with high and sharp spectral weight on the FS, indicating that this is a normal metal phase: the area enclosed by the FS equals to one-half of the Brillouin Zone, which satisfies the Luttinger's theorem as the fermion filling is fixed as one per site.
As $h$ increases, the spectral weight on the FS decreases, and vanishes at the critical value $h_c\sim2$ [Fig.~\ref{fig:fig2} (b)].
This is also reflected in the plot of the spectral weight in Fig.~\ref{fig:fig4} (a), as will be elucidated later. In particular, the evolution of the FS shows no sign of any symmetry-breaking across the transition  (change of the shape of the FS, if it were to be consistent with Luttinger's theorem), it is only the spectral weight along the FS vanishes.
Finally, for $h>h_c$, the FS completely disappears [Fig.~\ref{fig:fig2} (c)]. In this phase, the system behaves like an insulator from a spectral perspective, if one were able to perform ARPES experiment on the OM phase, the experimentalist will detect an insulator with single particle gap.

The absence of the FS appears to violate the Luttinger's theorem (note that the fermion filling is still fixed at one by the particle-hole symmetry), unless the volumn of a hidden $f$-fermion FS is also included. Such an exotic phase state, which actually has metallic responses from other perspectives as will be revealed below, is an OM beyond existing paradigm of metals, and it is the major discovery of this work.

Next, we examine the OM more closely. Despite of the absence of FS, as indicated by the upper panels of Fig.~\ref{fig:fig2}, there is still a hidden FS, which is associated with the $f$-fermions, as required by the Luttinger's theorem.
This can be observed through the magnetic response. Fig.~\ref{fig:fig2} (d), (e) and (f) show the magnetic response of the system across the NM-to-OM transition. What is calculated is the magnetic susceptibility of the $c$ fermions, $\chi(\mathbf{q},\omega=0)=\frac{1}{\beta N}\int^{\beta}_{0}d\tau \sum_{i,j} e^{i\mathbf{q}\cdot \mathbf{r}_{ij}}\langle (n^{\uparrow}_{i,c}-n^{\downarrow}_{i,c})(\tau)(n^{\uparrow}_{j,c}-n^{\downarrow}_{j,c})(0)\rangle$.
This quantity is gauge-neutral and demonstrates the magnetic response of the system (notice that $\chi$ is also the magnetic susceptibility of the orthogonal fermion $f$, as $c$ is related to $f$ by $c_{i\alpha}=f_{i\alpha}S_i^z$, and $(S_i^z)^2=+1$.)
It is very interesting to see that there is little change in the $\chi(\mathbf{q},\omega=0)$ across the NM-to-OM transition, from Fig.~\ref{fig:fig2} (d) to (f).
Inside the NM phase, the strongest magnetic responses are at $\mathbf{q}=(\pi,\pi)$, which stems from the nesting of the dimond-shape FS in Fig.~\ref{fig:fig2} (a), such a FS having antiferromagnetic instability is well-known and investigated~\cite{Hirsch1985}. Hence, the $(\pi,\pi)$ peak in the NM phase can be attributed to the $c$-fermion FS.
However, close to and inside the OM phase [Fig.~\ref{fig:fig2} (e) and (f)], despite of the absence of FS, the $\chi(\mathbf{q},\omega=0)$ still peaks at $\mathbf{q}=(\pi,\pi)$ and the amplitude of the susceptibility is almost unchanged, this is further illustrated in Fig.~\ref{fig:fig4} (b) as a function of $h$, will be elucidated later. This means that there still exists a FS structure comprised of the $f$ fermions, whose magnetic response resembles that of a non-interacting half-filled square-lattice fermion model. Although the OM is a strongly interacting phase, in which the $f$ fermion, Ising matter field $S^{z}$ and the $\mathbb{Z}_2$ gauge field $\sigma^{z}$ are strongly coupled together, and that it wouldn't respond to the single-particle spectral probe, but Fig.~\ref{fig:fig2} (e) and (f) show that the OM is metal in disguise and has the same magnetic responses as that of the NM phase with FS instabilities.

Next we turn to the continuous quantum phase transition between the NM and the OM phases. Since there is no local order parameter associated with this transition, we cannot performed the usual finite size analysis based on the correlation functions associated with the order parameter for this transition. Therefore, to determine the precise position of the QCP, one can monitor the energy derivative and the second derivative over the control parameter $h$, which is shown in Fig.~\ref{fig:fig3} (a) and (b), this is also a common measurement for detecting the position of the transition~\cite{YYHe2016,XYXuSO42017}. $\partial \langle E \rangle / \partial h = \frac{1}{N}\sum_{i}\langle S^{x}_{i} \rangle$ serves as the first order derivative of the internal energy of the system over the control parameter of the transition~\cite{YYHe2016}, and a change of the slope can be seen at $h=h_c = 2.2(2)$ (highlighted by the vertical dash line).  Since the first order derivative of the internal energy is still continuous, that NM-to-OM transition shall be a continuous transition as well. The position of the transition is more obvious in the second derivative in Fig.~\ref{fig:fig3} (b), with three different system sizes $L=12,14$ and $16$, the change of the slope in Fig.~\ref{fig:fig3} (a) manifests as a clear peak here, and the position of the peak, highlighted by the vertical dashed line, is again at $h_c=2.2$.

As discussed in previous section, this transition is a Higgs transition, because the order parameter $S_i^z$ carries a nontrivial $\mathbb Z_2$ charge.
As a consequence, the phase transition cannot be observed by directly measuring the correlation function of $\langle S_i^zS_{i+\mathbf{r}}^z\rangle$. (In a gauge theory, all non-gauge-invariant correlation functions vanish~\cite{Kogut1979}.)
Instead, this Higgs transition can be detected by constructing a string operator, $C(\mathbf{r})=\langle S^{z}_{i}\prod_{i}^{i+\mathbf{r}}\sigma^{z}_{b\in \hat{x}} S^{z}_{i+\mathbf{r}} \rangle$, which measures the gauge-invariant correlation function of the Ising matter field attached with string of $\mathbb{Z}_2$ gauge fields connecting the sites $i$ and $i+\mathbf{r}$. The results are shown in Fig.~\ref{fig:fig3} (c). For the sake of simplicity, we have chosen a path along the $\hat{x}$ direction of the lattice. It is clear that at $h<h_c$ (the blue curve of $h=0.4$), when the system is inside the NM phase, the string operator demonstrates a hidden long-range order, although there is no long-range order of any local operator and thus no real symmetry-breaking in the system. As $h$ gradually increases towards $h_c$, the long-range order in $C(x)$ becomes weaker (such as the red curve for $h=2.0$), and when $h>h_c$ at $h=4.0$, the correlation is completely short-ranged, meaning that inside the OM phase there is no long-range order in the string order.
The observation of the string order parameter confirms our understanding of the two phases:
In the OM phase, the $\mathbb{Z}_2$ gauge field is deconfined~\cite{Assaad2016,Gazit2018,Gazit2019}, which comes hand-in-hand with the vanishing FS of the $c$ fermions, hidden FS of the $f$ fermions and appearing of $\mathbb{Z}_2$ topological order of the $\sigma^{z}$ gauge field.
In the NM phase, the ordering of $C(\mathbf{r})$ has two consequences: $\mathbb Z_2$ topological order disappears ($\mathbb Z_2$ gauge field is ``Higgsed''), and the hidden $f$-fermion FS becomes a $c$-fermion FS because $c$ and $f$ fermions can be identified as $c_i=f_iS_i^z$.
Due to its topological and interacting nature, the quantum critical properties of the OM-NM transition such as its field theory description and exponents will be of highly theoretical interests and will be addressed in future studies.

\section{Discussion}

The quantum phase transition from NM to OM is highly non-trivial as it is where fractionalization, dynamical gauge fluctuations, a hidden FS and a topological order all come together. The more detailed information at and across the transition will be helpful for the further development of the theoretical description of this QCP. Fig.~\ref{fig:fig4} (a) provides spectral weight/single-particle residual, $Z(\mathbf{k}_F)\sim G(\mathbf{k}_F,\beta/2)$, at two different momenta across the transition. It is clear that for both $\mathbf{k}_F = (\pi,0)$ and $(\pi/2,\pi/2)$, single-particle residual continuously reduces to zero at $h_c$. Deep inside the NM ($h<h_c$), $Z(\mathbf{k}_{F})\approx 1$, signifying FL nature of the phase; deep inside the OM ($h>h_c$), $Z(\mathbf{k}_F) \approx 0$, signifying the nFL nature of the phase. The associated magnetic susceptibility, $\chi(\mathbf{q},\omega=0)$ with $\mathbf{q}=(\pi,\pi)$, is shown in Fig.~\ref{fig:fig4} (b) as a function of $h$. As discussed in the Fig.~\ref{fig:fig2}, the $\chi(\mathbf{q})$ does not develop any singularity across $h_c$, instead, it is kept almost a constant throughout. In the meantime, Fig.~\ref{fig:fig4} (c) depicts the average fermion density $\langle n\rangle$ as a function of chemical potential, for $h=4$ inside the OM phase. The plot shows a continuous curve with no incompressible region, i.e. no flat segment in $\langle n\rangle(\mu)$, which further support that the OM phase is not a charge insulator. This, combined the Fig.~\ref{fig:fig4} (b), are intriguing in that both NM and OM behave in the same way in magnetic and charge response, yet NM is a metal from single-particle spectrum while OM is an insulator in that respect.

The discovery of the OM phase and the apparently continuous quantum phase transition between NM and OM, paves the way to further investigate quantum metals that are beyond Landau's Fermi liquid paradigm. What is unique in our finding here is that the lattice model in Eq.~\eqref{eq:H} is solved in an unambiguous manner with QMC, which solidifies the existence of OM phase and its QCP with NM. Together with works such as Ref.~\cite{Hohenadler2018,Hohenadler2019} where a similar OM state is discovered at finite temperature and Refs.~\cite{Assaad2016,Gazit2016,Gazit2018,XiaoYanXu2018} where the deconfinement-confinement transition of the Dirac fermions are revealed and Ref.~\cite{Gazit2019} where an orthogonal semi-metal is discoveried, our results have now completed the models upon which the new paradigm of quantum metals can be firmly established. That is, without symmetry-breaking of any type, a FL can go through a QCP to a OM with no quasiparticle fraction but still responds towards other perturbations, just like a metal, due to the existence of a hidden FS with fractionalized fermionic degrees of freedom carrying charges and spins. It is not immediately clear that our findings could explain the perplexing experimental observation in nFLs such as those in the pseudogap in high-Tc superconductors, heavy fermion compounds, etc. But it is clear that since the awkward cousin of metal is finalized found, given time and patience, we will be able to know he/she better and could hope to eventually make the acquaintance with the entire family, in which more interesting characters are awaiting.

\section*{acknowledgments}
We thank Subhro Bhattacharjee, Snir Gazit, Fakher Assaad, Max Metlitski, Todadri Senthil, Subir Sachdev and Anders Sandvik for helpful discussions. CC and ZYM acknowledge the supports from the Ministry of Science and Technology of China through the National Key Research and Development Program (Grant No. 2016YFA0300502), the National Science Foundation of China (Grant No. 11574359 and 11674370) and the Research Grants Council of Hong Kong SAR China through 17303019. X. Y. X. is thankful for the support of the Research Grants Council of Hong Kong SAR China through C6026-16W, 16324216 and 16307117. YQ acknowledges supports from Minstry of Science and Technology of China under Grant No. 2015CB921700, and from National Science Foundation of China under Grant No. 11874115. We thank the Center for Quantum Simulation Sciences in the Institute of Physics, Chinese Academy of Sciences, the Computational Initiative at the Faculty of Science at the University of Hong Kong, the Platform for Data-Driven Computational Materials Discovery at the Songshan Lake Materials Laboratory, Guangdong, China and the Tianhe-1A platform at the National Supercomputer Center in Tianjin and Tianhe-2 platform at the National Supercomputer Center in Guangzhou for their technical support and generous allocation of CPU time.


\bibliographystyle{apsrev4-1}
\bibliography{main}

\begin{thebibliography}{46}%
\makeatletter
\providecommand \@ifxundefined [1]{%
 \@ifx{#1\undefined}
}%
\providecommand \@ifnum [1]{%
 \ifnum #1\expandafter \@firstoftwo
 \else \expandafter \@secondoftwo
 \fi
}%
\providecommand \@ifx [1]{%
 \ifx #1\expandafter \@firstoftwo
 \else \expandafter \@secondoftwo
 \fi
}%
\providecommand \natexlab [1]{#1}%
\providecommand \enquote  [1]{``#1''}%
\providecommand \bibnamefont  [1]{#1}%
\providecommand \bibfnamefont [1]{#1}%
\providecommand \citenamefont [1]{#1}%
\providecommand \href@noop [0]{\@secondoftwo}%
\providecommand \href [0]{\begingroup \@sanitize@url \@href}%
\providecommand \@href[1]{\@@startlink{#1}\@@href}%
\providecommand \@@href[1]{\endgroup#1\@@endlink}%
\providecommand \@sanitize@url [0]{\catcode `\\12\catcode `\$12\catcode
  `\&12\catcode `\#12\catcode `\^12\catcode `\_12\catcode `\%12\relax}%
\providecommand \@@startlink[1]{}%
\providecommand \@@endlink[0]{}%
\providecommand \url  [0]{\begingroup\@sanitize@url \@url }%
\providecommand \@url [1]{\endgroup\@href {#1}{\urlprefix }}%
\providecommand \urlprefix  [0]{URL }%
\providecommand \Eprint [0]{\href }%
\providecommand \doibase [0]{http://dx.doi.org/}%
\providecommand \selectlanguage [0]{\@gobble}%
\providecommand \bibinfo  [0]{\@secondoftwo}%
\providecommand \bibfield  [0]{\@secondoftwo}%
\providecommand \translation [1]{[#1]}%
\providecommand \BibitemOpen [0]{}%
\providecommand \bibitemStop [0]{}%
\providecommand \bibitemNoStop [0]{.\EOS\space}%
\providecommand \EOS [0]{\spacefactor3000\relax}%
\providecommand \BibitemShut  [1]{\csname bibitem#1\endcsname}%
\let\auto@bib@innerbib\@empty
\bibitem [{\citenamefont {Nandkishore}\ \emph {et~al.}(2012)\citenamefont
  {Nandkishore}, \citenamefont {Metlitski},\ and\ \citenamefont
  {Senthil}}]{Nandkishore2012}%
  \BibitemOpen
  \bibfield  {author} {\bibinfo {author} {\bibfnamefont {R.}~\bibnamefont
  {Nandkishore}}, \bibinfo {author} {\bibfnamefont {M.~A.}\ \bibnamefont
  {Metlitski}}, \ and\ \bibinfo {author} {\bibfnamefont {T.}~\bibnamefont
  {Senthil}},\ }\href {\doibase 10.1103/PhysRevB.86.045128} {\bibfield
  {journal} {\bibinfo  {journal} {Phys. Rev. B}\ }\textbf {\bibinfo {volume}
  {86}},\ \bibinfo {pages} {045128} (\bibinfo {year} {2012})}\BibitemShut
  {NoStop}%
\bibitem [{\citenamefont {Kaul}(2012)}]{Kaul2012}%
  \BibitemOpen
  \bibfield  {author} {\bibinfo {author} {\bibfnamefont {R.~K.}\ \bibnamefont
  {Kaul}},\ }\href {https://physics.aps.org/articles/v5/82} {\bibfield
  {journal} {\bibinfo  {journal} {Physics}\ }\textbf {\bibinfo {volume} {5}},\
  \bibinfo {pages} {82} (\bibinfo {year} {2012})}\BibitemShut {NoStop}%
\bibitem [{\citenamefont {R\"uegg}\ \emph {et~al.}(2010)\citenamefont
  {R\"uegg}, \citenamefont {Huber},\ and\ \citenamefont
  {Sigrist}}]{Reuegg2010}%
  \BibitemOpen
  \bibfield  {author} {\bibinfo {author} {\bibfnamefont {A.}~\bibnamefont
  {R\"uegg}}, \bibinfo {author} {\bibfnamefont {S.~D.}\ \bibnamefont {Huber}},
  \ and\ \bibinfo {author} {\bibfnamefont {M.}~\bibnamefont {Sigrist}},\ }\href
  {\doibase 10.1103/PhysRevB.81.155118} {\bibfield  {journal} {\bibinfo
  {journal} {Phys. Rev. B}\ }\textbf {\bibinfo {volume} {81}},\ \bibinfo
  {pages} {155118} (\bibinfo {year} {2010})}\BibitemShut {NoStop}%
\bibitem [{\citenamefont {L\"ohneysen}\ \emph {et~al.}(2007)\citenamefont
  {L\"ohneysen}, \citenamefont {Rosch}, \citenamefont {Vojta},\ and\
  \citenamefont {W\"olfle}}]{Loehneysen2007}%
  \BibitemOpen
  \bibfield  {author} {\bibinfo {author} {\bibfnamefont {H.~v.}\ \bibnamefont
  {L\"ohneysen}}, \bibinfo {author} {\bibfnamefont {A.}~\bibnamefont {Rosch}},
  \bibinfo {author} {\bibfnamefont {M.}~\bibnamefont {Vojta}}, \ and\ \bibinfo
  {author} {\bibfnamefont {P.}~\bibnamefont {W\"olfle}},\ }\href {\doibase
  10.1103/RevModPhys.79.1015} {\bibfield  {journal} {\bibinfo  {journal} {Rev.
  Mod. Phys.}\ }\textbf {\bibinfo {volume} {79}},\ \bibinfo {pages} {1015}
  (\bibinfo {year} {2007})}\BibitemShut {NoStop}%
\bibitem [{\citenamefont {Keimer}\ \emph {et~al.}(2015)\citenamefont {Keimer},
  \citenamefont {Kivelson}, \citenamefont {Norman}, \citenamefont {Uchida},\
  and\ \citenamefont {Zaanen}}]{Keimer2015}%
  \BibitemOpen
  \bibfield  {author} {\bibinfo {author} {\bibfnamefont {B.}~\bibnamefont
  {Keimer}}, \bibinfo {author} {\bibfnamefont {S.~A.}\ \bibnamefont
  {Kivelson}}, \bibinfo {author} {\bibfnamefont {M.~R.}\ \bibnamefont
  {Norman}}, \bibinfo {author} {\bibfnamefont {S.}~\bibnamefont {Uchida}}, \
  and\ \bibinfo {author} {\bibfnamefont {J.}~\bibnamefont {Zaanen}},\ }\href
  {\doibase 10.1038/nature14165} {\bibfield  {journal} {\bibinfo  {journal}
  {Nature}\ }\textbf {\bibinfo {volume} {518}} (\bibinfo {year} {2015}),\
  10.1038/nature14165}\BibitemShut {NoStop}%
\bibitem [{\citenamefont {Badoux}\ \emph {et~al.}(2016)\citenamefont {Badoux},
  \citenamefont {Tabis}, \citenamefont {Laliberté}, \citenamefont
  {Grissonnanche}, \citenamefont {Vignolle}, \citenamefont {Vignolles},
  \citenamefont {Béard}, \citenamefont {Bonn}, \citenamefont {Hardy},
  \citenamefont {Liang}, \citenamefont {Doiron-Leyraud}, \citenamefont
  {Taillefer},\ and\ \citenamefont {Proust}}]{Badoux2016}%
  \BibitemOpen
  \bibfield  {author} {\bibinfo {author} {\bibfnamefont {S.}~\bibnamefont
  {Badoux}}, \bibinfo {author} {\bibfnamefont {W.}~\bibnamefont {Tabis}},
  \bibinfo {author} {\bibfnamefont {F.}~\bibnamefont {Laliberté}}, \bibinfo
  {author} {\bibfnamefont {G.}~\bibnamefont {Grissonnanche}}, \bibinfo {author}
  {\bibfnamefont {B.}~\bibnamefont {Vignolle}}, \bibinfo {author}
  {\bibfnamefont {D.}~\bibnamefont {Vignolles}}, \bibinfo {author}
  {\bibfnamefont {J.}~\bibnamefont {Béard}}, \bibinfo {author} {\bibfnamefont
  {D.~A.}\ \bibnamefont {Bonn}}, \bibinfo {author} {\bibfnamefont {W.~N.}\
  \bibnamefont {Hardy}}, \bibinfo {author} {\bibfnamefont {R.}~\bibnamefont
  {Liang}}, \bibinfo {author} {\bibfnamefont {N.}~\bibnamefont
  {Doiron-Leyraud}}, \bibinfo {author} {\bibfnamefont {L.}~\bibnamefont
  {Taillefer}}, \ and\ \bibinfo {author} {\bibfnamefont {C.}~\bibnamefont
  {Proust}},\ }\href {\doibase 10.1038/nature16983} {\bibfield  {journal}
  {\bibinfo  {journal} {Nature}\ }\textbf {\bibinfo {volume} {531}},\ \bibinfo
  {pages} {210} (\bibinfo {year} {2016})}\BibitemShut {NoStop}%
\bibitem [{\citenamefont {Gu}\ \emph {et~al.}(2017)\citenamefont {Gu},
  \citenamefont {Liu}, \citenamefont {Xie}, \citenamefont {Zhang},
  \citenamefont {Gong}, \citenamefont {Hu}, \citenamefont {Ma}, \citenamefont
  {Li}, \citenamefont {Zhao}, \citenamefont {Lin}, \citenamefont {Xu},
  \citenamefont {Tan}, \citenamefont {Chen}, \citenamefont {Meng},
  \citenamefont {Yang}, \citenamefont {Luo},\ and\ \citenamefont
  {Li}}]{YHGu2017}%
  \BibitemOpen
  \bibfield  {author} {\bibinfo {author} {\bibfnamefont {Y.}~\bibnamefont
  {Gu}}, \bibinfo {author} {\bibfnamefont {Z.}~\bibnamefont {Liu}}, \bibinfo
  {author} {\bibfnamefont {T.}~\bibnamefont {Xie}}, \bibinfo {author}
  {\bibfnamefont {W.}~\bibnamefont {Zhang}}, \bibinfo {author} {\bibfnamefont
  {D.}~\bibnamefont {Gong}}, \bibinfo {author} {\bibfnamefont {D.}~\bibnamefont
  {Hu}}, \bibinfo {author} {\bibfnamefont {X.}~\bibnamefont {Ma}}, \bibinfo
  {author} {\bibfnamefont {C.}~\bibnamefont {Li}}, \bibinfo {author}
  {\bibfnamefont {L.}~\bibnamefont {Zhao}}, \bibinfo {author} {\bibfnamefont
  {L.}~\bibnamefont {Lin}}, \bibinfo {author} {\bibfnamefont {Z.}~\bibnamefont
  {Xu}}, \bibinfo {author} {\bibfnamefont {G.}~\bibnamefont {Tan}}, \bibinfo
  {author} {\bibfnamefont {G.}~\bibnamefont {Chen}}, \bibinfo {author}
  {\bibfnamefont {Z.~Y.}\ \bibnamefont {Meng}}, \bibinfo {author}
  {\bibfnamefont {Y.-f.}\ \bibnamefont {Yang}}, \bibinfo {author}
  {\bibfnamefont {H.}~\bibnamefont {Luo}}, \ and\ \bibinfo {author}
  {\bibfnamefont {S.}~\bibnamefont {Li}},\ }\href {\doibase
  10.1103/PhysRevLett.119.157001} {\bibfield  {journal} {\bibinfo  {journal}
  {Phys. Rev. Lett.}\ }\textbf {\bibinfo {volume} {119}},\ \bibinfo {pages}
  {157001} (\bibinfo {year} {2017})}\BibitemShut {NoStop}%
\bibitem [{\citenamefont {Stewart}(2001)}]{Stewart2001}%
  \BibitemOpen
  \bibfield  {author} {\bibinfo {author} {\bibfnamefont {G.~R.}\ \bibnamefont
  {Stewart}},\ }\href {\doibase 10.1103/RevModPhys.73.797} {\bibfield
  {journal} {\bibinfo  {journal} {Rev. Mod. Phys.}\ }\textbf {\bibinfo {volume}
  {73}},\ \bibinfo {pages} {797} (\bibinfo {year} {2001})}\BibitemShut
  {NoStop}%
\bibitem [{\citenamefont {Custers}\ \emph {et~al.}(2003)\citenamefont
  {Custers}, \citenamefont {Gegenwart}, \citenamefont {Wilhelm}, \citenamefont
  {Neumaier}, \citenamefont {Tokiwa}, \citenamefont {Trovarelli}, \citenamefont
  {Geibel}, \citenamefont {Steglich}, \citenamefont {Pepin},\ and\
  \citenamefont {Coleman}}]{Custers2003}%
  \BibitemOpen
  \bibfield  {author} {\bibinfo {author} {\bibfnamefont {J.}~\bibnamefont
  {Custers}}, \bibinfo {author} {\bibfnamefont {P.}~\bibnamefont {Gegenwart}},
  \bibinfo {author} {\bibfnamefont {H.}~\bibnamefont {Wilhelm}}, \bibinfo
  {author} {\bibfnamefont {K.}~\bibnamefont {Neumaier}}, \bibinfo {author}
  {\bibfnamefont {Y.}~\bibnamefont {Tokiwa}}, \bibinfo {author} {\bibfnamefont
  {O.}~\bibnamefont {Trovarelli}}, \bibinfo {author} {\bibfnamefont
  {C.}~\bibnamefont {Geibel}}, \bibinfo {author} {\bibfnamefont
  {F.}~\bibnamefont {Steglich}}, \bibinfo {author} {\bibfnamefont
  {C.}~\bibnamefont {Pepin}}, \ and\ \bibinfo {author} {\bibfnamefont
  {P.}~\bibnamefont {Coleman}},\ }\href {http://dx.doi.org/10.1038/nature01774}
  {\bibfield  {journal} {\bibinfo  {journal} {Nature}\ }\textbf {\bibinfo
  {volume} {424}},\ \bibinfo {pages} {524 } (\bibinfo {year}
  {2003})}\BibitemShut {NoStop}%
\bibitem [{\citenamefont {Si}\ and\ \citenamefont {Steglich}(2010)}]{QMSi2010}%
  \BibitemOpen
  \bibfield  {author} {\bibinfo {author} {\bibfnamefont {Q.}~\bibnamefont
  {Si}}\ and\ \bibinfo {author} {\bibfnamefont {F.}~\bibnamefont {Steglich}},\
  }\href {\doibase 10.1126/science.1191195} {\bibfield  {journal} {\bibinfo
  {journal} {Science}\ }\textbf {\bibinfo {volume} {329}},\ \bibinfo {pages}
  {1161} (\bibinfo {year} {2010})}\BibitemShut {NoStop}%
\bibitem [{\citenamefont {Steppke}\ \emph {et~al.}(2013)\citenamefont
  {Steppke}, \citenamefont {K{\"u}chler}, \citenamefont {Lausberg},
  \citenamefont {Lengyel}, \citenamefont {Steinke}, \citenamefont {Borth},
  \citenamefont {L{\"u}hmann}, \citenamefont {Krellner}, \citenamefont
  {Nicklas}, \citenamefont {Geibel}, \citenamefont {Steglich},\ and\
  \citenamefont {Brando}}]{Steppke2013}%
  \BibitemOpen
  \bibfield  {author} {\bibinfo {author} {\bibfnamefont {A.}~\bibnamefont
  {Steppke}}, \bibinfo {author} {\bibfnamefont {R.}~\bibnamefont
  {K{\"u}chler}}, \bibinfo {author} {\bibfnamefont {S.}~\bibnamefont
  {Lausberg}}, \bibinfo {author} {\bibfnamefont {E.}~\bibnamefont {Lengyel}},
  \bibinfo {author} {\bibfnamefont {L.}~\bibnamefont {Steinke}}, \bibinfo
  {author} {\bibfnamefont {R.}~\bibnamefont {Borth}}, \bibinfo {author}
  {\bibfnamefont {T.}~\bibnamefont {L{\"u}hmann}}, \bibinfo {author}
  {\bibfnamefont {C.}~\bibnamefont {Krellner}}, \bibinfo {author}
  {\bibfnamefont {M.}~\bibnamefont {Nicklas}}, \bibinfo {author} {\bibfnamefont
  {C.}~\bibnamefont {Geibel}}, \bibinfo {author} {\bibfnamefont
  {F.}~\bibnamefont {Steglich}}, \ and\ \bibinfo {author} {\bibfnamefont
  {M.}~\bibnamefont {Brando}},\ }\href {\doibase 10.1126/science.1230583}
  {\bibfield  {journal} {\bibinfo  {journal} {Science}\ }\textbf {\bibinfo
  {volume} {339}},\ \bibinfo {pages} {933} (\bibinfo {year}
  {2013})}\BibitemShut {NoStop}%
\bibitem [{\citenamefont {Cao}\ \emph {et~al.}(2018{\natexlab{a}})\citenamefont
  {Cao}, \citenamefont {Fatemi}, \citenamefont {Demir}, \citenamefont {Fang},
  \citenamefont {Tomarken}, \citenamefont {Luo}, \citenamefont
  {Sanchez-Yamagishi}, \citenamefont {Watanabe}, \citenamefont {Taniguchi},
  \citenamefont {Kaxiras}, \citenamefont {Ashoori},\ and\ \citenamefont
  {Jarillo-Herrero}}]{cao2018correlated}%
  \BibitemOpen
  \bibfield  {author} {\bibinfo {author} {\bibfnamefont {Y.}~\bibnamefont
  {Cao}}, \bibinfo {author} {\bibfnamefont {V.}~\bibnamefont {Fatemi}},
  \bibinfo {author} {\bibfnamefont {A.}~\bibnamefont {Demir}}, \bibinfo
  {author} {\bibfnamefont {S.}~\bibnamefont {Fang}}, \bibinfo {author}
  {\bibfnamefont {S.~L.}\ \bibnamefont {Tomarken}}, \bibinfo {author}
  {\bibfnamefont {J.~Y.}\ \bibnamefont {Luo}}, \bibinfo {author} {\bibfnamefont
  {J.~D.}\ \bibnamefont {Sanchez-Yamagishi}}, \bibinfo {author} {\bibfnamefont
  {K.}~\bibnamefont {Watanabe}}, \bibinfo {author} {\bibfnamefont
  {T.}~\bibnamefont {Taniguchi}}, \bibinfo {author} {\bibfnamefont
  {E.}~\bibnamefont {Kaxiras}}, \bibinfo {author} {\bibfnamefont {R.~C.}\
  \bibnamefont {Ashoori}}, \ and\ \bibinfo {author} {\bibfnamefont
  {P.}~\bibnamefont {Jarillo-Herrero}},\ }\href {\doibase 10.1038/nature26154}
  {\bibfield  {journal} {\bibinfo  {journal} {Nature}\ }\textbf {\bibinfo
  {volume} {556}},\ \bibinfo {pages} {80} (\bibinfo {year}
  {2018}{\natexlab{a}})}\BibitemShut {NoStop}%
\bibitem [{\citenamefont {Cao}\ \emph {et~al.}(2018{\natexlab{b}})\citenamefont
  {Cao}, \citenamefont {Fatemi}, \citenamefont {Fang}, \citenamefont
  {Watanabe}, \citenamefont {Taniguchi}, \citenamefont {Kaxiras},\ and\
  \citenamefont {Jarillo-Herrero}}]{cao2018unconventional}%
  \BibitemOpen
  \bibfield  {author} {\bibinfo {author} {\bibfnamefont {Y.}~\bibnamefont
  {Cao}}, \bibinfo {author} {\bibfnamefont {V.}~\bibnamefont {Fatemi}},
  \bibinfo {author} {\bibfnamefont {S.}~\bibnamefont {Fang}}, \bibinfo {author}
  {\bibfnamefont {K.}~\bibnamefont {Watanabe}}, \bibinfo {author}
  {\bibfnamefont {T.}~\bibnamefont {Taniguchi}}, \bibinfo {author}
  {\bibfnamefont {E.}~\bibnamefont {Kaxiras}}, \ and\ \bibinfo {author}
  {\bibfnamefont {P.}~\bibnamefont {Jarillo-Herrero}},\ }\href {\doibase
  10.1038/nature26160} {\bibfield  {journal} {\bibinfo  {journal} {Nature}\
  }\textbf {\bibinfo {volume} {556}},\ \bibinfo {pages} {43} (\bibinfo {year}
  {2018}{\natexlab{b}})}\BibitemShut {NoStop}%
\bibitem [{\citenamefont {{Cao}}\ \emph {et~al.}(2019)\citenamefont {{Cao}},
  \citenamefont {{Chowdhury}}, \citenamefont {{Rodan-Legrain}}, \citenamefont
  {{Rubies-Bigord{\`a}}}, \citenamefont {{Watanabe}}, \citenamefont
  {{Taniguchi}}, \citenamefont {{Senthil}},\ and\ \citenamefont
  {{Jarillo-Herrero}}}]{YuanCao2019}%
  \BibitemOpen
  \bibfield  {author} {\bibinfo {author} {\bibfnamefont {Y.}~\bibnamefont
  {{Cao}}}, \bibinfo {author} {\bibfnamefont {D.}~\bibnamefont {{Chowdhury}}},
  \bibinfo {author} {\bibfnamefont {D.}~\bibnamefont {{Rodan-Legrain}}},
  \bibinfo {author} {\bibfnamefont {O.}~\bibnamefont {{Rubies-Bigord{\`a}}}},
  \bibinfo {author} {\bibfnamefont {K.}~\bibnamefont {{Watanabe}}}, \bibinfo
  {author} {\bibfnamefont {T.}~\bibnamefont {{Taniguchi}}}, \bibinfo {author}
  {\bibfnamefont {T.}~\bibnamefont {{Senthil}}}, \ and\ \bibinfo {author}
  {\bibfnamefont {P.}~\bibnamefont {{Jarillo-Herrero}}},\ }\href@noop {}
  {\bibfield  {journal} {\bibinfo  {journal} {arXiv e-prints}\ ,\ \bibinfo
  {eid} {arXiv:1901.03710}} (\bibinfo {year} {2019})},\ \Eprint
  {http://arxiv.org/abs/1901.03710} {arXiv:1901.03710 [cond-mat.str-el]}
  \BibitemShut {NoStop}%
\bibitem [{\citenamefont {{Shen}}\ \emph {et~al.}(2019)\citenamefont {{Shen}},
  \citenamefont {{Li}}, \citenamefont {{Wang}}, \citenamefont {{Zhao}},
  \citenamefont {{Tang}}, \citenamefont {{Liu}}, \citenamefont {{Tian}},
  \citenamefont {{Chu}}, \citenamefont {{Watanabe}}, \citenamefont
  {{Taniguchi}}, \citenamefont {{Yang}}, \citenamefont {{Meng}}, \citenamefont
  {{Shi}},\ and\ \citenamefont {{Zhang}}}]{ChengShen2019}%
  \BibitemOpen
  \bibfield  {author} {\bibinfo {author} {\bibfnamefont {C.}~\bibnamefont
  {{Shen}}}, \bibinfo {author} {\bibfnamefont {N.}~\bibnamefont {{Li}}},
  \bibinfo {author} {\bibfnamefont {S.}~\bibnamefont {{Wang}}}, \bibinfo
  {author} {\bibfnamefont {Y.}~\bibnamefont {{Zhao}}}, \bibinfo {author}
  {\bibfnamefont {J.}~\bibnamefont {{Tang}}}, \bibinfo {author} {\bibfnamefont
  {J.}~\bibnamefont {{Liu}}}, \bibinfo {author} {\bibfnamefont
  {J.}~\bibnamefont {{Tian}}}, \bibinfo {author} {\bibfnamefont
  {Y.}~\bibnamefont {{Chu}}}, \bibinfo {author} {\bibfnamefont
  {K.}~\bibnamefont {{Watanabe}}}, \bibinfo {author} {\bibfnamefont
  {T.}~\bibnamefont {{Taniguchi}}}, \bibinfo {author} {\bibfnamefont
  {R.}~\bibnamefont {{Yang}}}, \bibinfo {author} {\bibfnamefont {Z.~Y.}\
  \bibnamefont {{Meng}}}, \bibinfo {author} {\bibfnamefont {D.}~\bibnamefont
  {{Shi}}}, \ and\ \bibinfo {author} {\bibfnamefont {G.}~\bibnamefont
  {{Zhang}}},\ }\href@noop {} {\bibfield  {journal} {\bibinfo  {journal} {arXiv
  e-prints}\ ,\ \bibinfo {eid} {arXiv:1903.06952}} (\bibinfo {year} {2019})},\
  \Eprint {http://arxiv.org/abs/1903.06952} {arXiv:1903.06952
  [cond-mat.supr-con]} \BibitemShut {NoStop}%
\bibitem [{\citenamefont {Luttinger}(1960)}]{Luttinger1960}%
  \BibitemOpen
  \bibfield  {author} {\bibinfo {author} {\bibfnamefont {J.~M.}\ \bibnamefont
  {Luttinger}},\ }\href {\doibase 10.1103/PhysRev.119.1153} {\bibfield
  {journal} {\bibinfo  {journal} {Phys. Rev.}\ }\textbf {\bibinfo {volume}
  {119}},\ \bibinfo {pages} {1153} (\bibinfo {year} {1960})}\BibitemShut
  {NoStop}%
\bibitem [{\citenamefont {Oshikawa}(2000)}]{Oshikawa2000}%
  \BibitemOpen
  \bibfield  {author} {\bibinfo {author} {\bibfnamefont {M.}~\bibnamefont
  {Oshikawa}},\ }\href {\doibase 10.1103/PhysRevLett.84.3370} {\bibfield
  {journal} {\bibinfo  {journal} {Phys. Rev. Lett.}\ }\textbf {\bibinfo
  {volume} {84}},\ \bibinfo {pages} {3370} (\bibinfo {year}
  {2000})}\BibitemShut {NoStop}%
\bibitem [{\citenamefont {Paramekanti}\ and\ \citenamefont
  {Vishwanath}(2004)}]{Paramekanti2004}%
  \BibitemOpen
  \bibfield  {author} {\bibinfo {author} {\bibfnamefont {A.}~\bibnamefont
  {Paramekanti}}\ and\ \bibinfo {author} {\bibfnamefont {A.}~\bibnamefont
  {Vishwanath}},\ }\href {\doibase 10.1103/PhysRevB.70.245118} {\bibfield
  {journal} {\bibinfo  {journal} {Phys. Rev. B}\ }\textbf {\bibinfo {volume}
  {70}},\ \bibinfo {pages} {245118} (\bibinfo {year} {2004})}\BibitemShut
  {NoStop}%
\bibitem [{\citenamefont {Wu}\ \emph {et~al.}(2014)\citenamefont {Wu},
  \citenamefont {Cheng}, \citenamefont {Matsubayashi}, \citenamefont {Kong},
  \citenamefont {Lin}, \citenamefont {Jin}, \citenamefont {Wang}, \citenamefont
  {Uwatoko},\ and\ \citenamefont {Luo}}]{Wu2014}%
  \BibitemOpen
  \bibfield  {author} {\bibinfo {author} {\bibfnamefont {W.}~\bibnamefont
  {Wu}}, \bibinfo {author} {\bibfnamefont {J.}~\bibnamefont {Cheng}}, \bibinfo
  {author} {\bibfnamefont {K.}~\bibnamefont {Matsubayashi}}, \bibinfo {author}
  {\bibfnamefont {P.}~\bibnamefont {Kong}}, \bibinfo {author} {\bibfnamefont
  {F.}~\bibnamefont {Lin}}, \bibinfo {author} {\bibfnamefont {C.}~\bibnamefont
  {Jin}}, \bibinfo {author} {\bibfnamefont {N.}~\bibnamefont {Wang}}, \bibinfo
  {author} {\bibfnamefont {Y.}~\bibnamefont {Uwatoko}}, \ and\ \bibinfo
  {author} {\bibfnamefont {J.}~\bibnamefont {Luo}},\ }\href
  {http://dx.doi.org/10.1038/ncomms6508} {\bibfield  {journal} {\bibinfo
  {journal} {Nature Communications}\ }\textbf {\bibinfo {volume} {5}},\
  \bibinfo {pages} {5508} (\bibinfo {year} {2014})}\BibitemShut {NoStop}%
\bibitem [{\citenamefont {Cheng}\ and\ \citenamefont {Luo}(2017)}]{Cheng2017}%
  \BibitemOpen
  \bibfield  {author} {\bibinfo {author} {\bibfnamefont {J.}~\bibnamefont
  {Cheng}}\ and\ \bibinfo {author} {\bibfnamefont {J.}~\bibnamefont {Luo}},\
  }\href {http://stacks.iop.org/0953-8984/29/i=38/a=383003} {\bibfield
  {journal} {\bibinfo  {journal} {Journal of Physics: Condensed Matter}\
  }\textbf {\bibinfo {volume} {29}},\ \bibinfo {pages} {383003} (\bibinfo
  {year} {2017})}\BibitemShut {NoStop}%
\bibitem [{\citenamefont {Matsuda}\ \emph {et~al.}(2018)\citenamefont
  {Matsuda}, \citenamefont {Lin}, \citenamefont {Yu}, \citenamefont {Cheng},
  \citenamefont {Wu}, \citenamefont {Sun}, \citenamefont {Zhang}, \citenamefont
  {Sun}, \citenamefont {Matsubayashi}, \citenamefont {Miyake}, \citenamefont
  {Kato}, \citenamefont {Yan}, \citenamefont {Stone}, \citenamefont {Si},
  \citenamefont {Luo},\ and\ \citenamefont {Uwatoko}}]{JGCheng2018}%
  \BibitemOpen
  \bibfield  {author} {\bibinfo {author} {\bibfnamefont {M.}~\bibnamefont
  {Matsuda}}, \bibinfo {author} {\bibfnamefont {F.~K.}\ \bibnamefont {Lin}},
  \bibinfo {author} {\bibfnamefont {R.}~\bibnamefont {Yu}}, \bibinfo {author}
  {\bibfnamefont {J.-G.}\ \bibnamefont {Cheng}}, \bibinfo {author}
  {\bibfnamefont {W.}~\bibnamefont {Wu}}, \bibinfo {author} {\bibfnamefont
  {J.~P.}\ \bibnamefont {Sun}}, \bibinfo {author} {\bibfnamefont {J.~H.}\
  \bibnamefont {Zhang}}, \bibinfo {author} {\bibfnamefont {P.~J.}\ \bibnamefont
  {Sun}}, \bibinfo {author} {\bibfnamefont {K.}~\bibnamefont {Matsubayashi}},
  \bibinfo {author} {\bibfnamefont {T.}~\bibnamefont {Miyake}}, \bibinfo
  {author} {\bibfnamefont {T.}~\bibnamefont {Kato}}, \bibinfo {author}
  {\bibfnamefont {J.-Q.}\ \bibnamefont {Yan}}, \bibinfo {author} {\bibfnamefont
  {M.~B.}\ \bibnamefont {Stone}}, \bibinfo {author} {\bibfnamefont
  {Q.}~\bibnamefont {Si}}, \bibinfo {author} {\bibfnamefont {J.~L.}\
  \bibnamefont {Luo}}, \ and\ \bibinfo {author} {\bibfnamefont
  {Y.}~\bibnamefont {Uwatoko}},\ }\href {\doibase 10.1103/PhysRevX.8.031017}
  {\bibfield  {journal} {\bibinfo  {journal} {Phys. Rev. X}\ }\textbf {\bibinfo
  {volume} {8}},\ \bibinfo {pages} {031017} (\bibinfo {year}
  {2018})}\BibitemShut {NoStop}%
\bibitem [{\citenamefont {Senthil}\ \emph {et~al.}(2003)\citenamefont
  {Senthil}, \citenamefont {Sachdev},\ and\ \citenamefont
  {Vojta}}]{Senthil2003}%
  \BibitemOpen
  \bibfield  {author} {\bibinfo {author} {\bibfnamefont {T.}~\bibnamefont
  {Senthil}}, \bibinfo {author} {\bibfnamefont {S.}~\bibnamefont {Sachdev}}, \
  and\ \bibinfo {author} {\bibfnamefont {M.}~\bibnamefont {Vojta}},\ }\href
  {\doibase 10.1103/PhysRevLett.90.216403} {\bibfield  {journal} {\bibinfo
  {journal} {Phys. Rev. Lett.}\ }\textbf {\bibinfo {volume} {90}},\ \bibinfo
  {pages} {216403} (\bibinfo {year} {2003})}\BibitemShut {NoStop}%
\bibitem [{\citenamefont {Punk}\ \emph {et~al.}(2015)\citenamefont {Punk},
  \citenamefont {Allais},\ and\ \citenamefont {Sachdev}}]{Punk2015}%
  \BibitemOpen
  \bibfield  {author} {\bibinfo {author} {\bibfnamefont {M.}~\bibnamefont
  {Punk}}, \bibinfo {author} {\bibfnamefont {A.}~\bibnamefont {Allais}}, \ and\
  \bibinfo {author} {\bibfnamefont {S.}~\bibnamefont {Sachdev}},\ }\href
  {\doibase 10.1073/pnas.1512206112} {\bibfield  {journal} {\bibinfo  {journal}
  {Proceedings of the National Academy of Sciences}\ }\textbf {\bibinfo
  {volume} {112}},\ \bibinfo {pages} {9552} (\bibinfo {year}
  {2015})}\BibitemShut {NoStop}%
\bibitem [{\citenamefont {Feldmeier}\ \emph {et~al.}(2018)\citenamefont
  {Feldmeier}, \citenamefont {Huber},\ and\ \citenamefont
  {Punk}}]{Feldmeier2018}%
  \BibitemOpen
  \bibfield  {author} {\bibinfo {author} {\bibfnamefont {J.}~\bibnamefont
  {Feldmeier}}, \bibinfo {author} {\bibfnamefont {S.}~\bibnamefont {Huber}}, \
  and\ \bibinfo {author} {\bibfnamefont {M.}~\bibnamefont {Punk}},\ }\href
  {\doibase 10.1103/PhysRevLett.120.187001} {\bibfield  {journal} {\bibinfo
  {journal} {Phys. Rev. Lett.}\ }\textbf {\bibinfo {volume} {120}},\ \bibinfo
  {pages} {187001} (\bibinfo {year} {2018})}\BibitemShut {NoStop}%
\bibitem [{\citenamefont {Maldacena}\ and\ \citenamefont
  {Stanford}(2016)}]{MaldacenaStanford2016}%
  \BibitemOpen
  \bibfield  {author} {\bibinfo {author} {\bibfnamefont {J.}~\bibnamefont
  {Maldacena}}\ and\ \bibinfo {author} {\bibfnamefont {D.}~\bibnamefont
  {Stanford}},\ }\href {\doibase 10.1103/PhysRevD.94.106002} {\bibfield
  {journal} {\bibinfo  {journal} {Phys. Rev. D}\ }\textbf {\bibinfo {volume}
  {94}},\ \bibinfo {pages} {106002} (\bibinfo {year} {2016})}\BibitemShut
  {NoStop}%
\bibitem [{\citenamefont {Hofmann}\ \emph {et~al.}(2019)\citenamefont
  {Hofmann}, \citenamefont {Assaad},\ and\ \citenamefont
  {Grover}}]{Hofmann2018}%
  \BibitemOpen
  \bibfield  {author} {\bibinfo {author} {\bibfnamefont {J.~S.}\ \bibnamefont
  {Hofmann}}, \bibinfo {author} {\bibfnamefont {F.~F.}\ \bibnamefont {Assaad}},
  \ and\ \bibinfo {author} {\bibfnamefont {T.}~\bibnamefont {Grover}},\ }\href
  {\doibase 10.1103/PhysRevB.100.035118} {\bibfield  {journal} {\bibinfo
  {journal} {Phys. Rev. B}\ }\textbf {\bibinfo {volume} {100}},\ \bibinfo
  {pages} {035118} (\bibinfo {year} {2019})}\BibitemShut {NoStop}%
\bibitem [{\citenamefont {{Pan}}\ \emph {et~al.}(2020)\citenamefont {{Pan}},
  \citenamefont {{Wang}},\ and\ \citenamefont {{Meng}}}]{GPPan2020}%
  \BibitemOpen
  \bibfield  {author} {\bibinfo {author} {\bibfnamefont {G.}~\bibnamefont
  {{Pan}}}, \bibinfo {author} {\bibfnamefont {Y.}~\bibnamefont {{Wang}}}, \
  and\ \bibinfo {author} {\bibfnamefont {Z.~Y.}\ \bibnamefont {{Meng}}},\
  }\href@noop {} {\bibfield  {journal} {\bibinfo  {journal} {arXiv e-prints}\
  ,\ \bibinfo {eid} {arXiv:2001.06586}} (\bibinfo {year} {2020})},\ \Eprint
  {http://arxiv.org/abs/2001.06586} {arXiv:2001.06586 [cond-mat.str-el]}
  \BibitemShut {NoStop}%
\bibitem [{\citenamefont {{Gazit}}\ \emph {et~al.}(2019)\citenamefont
  {{Gazit}}, \citenamefont {{Assaad}},\ and\ \citenamefont
  {{Sachdev}}}]{Gazit2019}%
  \BibitemOpen
  \bibfield  {author} {\bibinfo {author} {\bibfnamefont {S.}~\bibnamefont
  {{Gazit}}}, \bibinfo {author} {\bibfnamefont {F.~F.}\ \bibnamefont
  {{Assaad}}}, \ and\ \bibinfo {author} {\bibfnamefont {S.}~\bibnamefont
  {{Sachdev}}},\ }\href@noop {} {\bibfield  {journal} {\bibinfo  {journal}
  {arXiv e-prints}\ ,\ \bibinfo {eid} {arXiv:1906.11250}} (\bibinfo {year}
  {2019})},\ \Eprint {http://arxiv.org/abs/1906.11250} {arXiv:1906.11250
  [cond-mat.str-el]} \BibitemShut {NoStop}%
\bibitem [{\citenamefont {Huijse}\ and\ \citenamefont
  {Sachdev}(2011)}]{Liza2011}%
  \BibitemOpen
  \bibfield  {author} {\bibinfo {author} {\bibfnamefont {L.}~\bibnamefont
  {Huijse}}\ and\ \bibinfo {author} {\bibfnamefont {S.}~\bibnamefont
  {Sachdev}},\ }\href {\doibase 10.1103/PhysRevD.84.026001} {\bibfield
  {journal} {\bibinfo  {journal} {Phys. Rev. D}\ }\textbf {\bibinfo {volume}
  {84}},\ \bibinfo {pages} {026001} (\bibinfo {year} {2011})}\BibitemShut
  {NoStop}%
\bibitem [{\citenamefont {Senthil}\ and\ \citenamefont
  {Motrunich}(2002)}]{Senthil2002}%
  \BibitemOpen
  \bibfield  {author} {\bibinfo {author} {\bibfnamefont {T.}~\bibnamefont
  {Senthil}}\ and\ \bibinfo {author} {\bibfnamefont {O.}~\bibnamefont
  {Motrunich}},\ }\href {\doibase 10.1103/PhysRevB.66.205104} {\bibfield
  {journal} {\bibinfo  {journal} {Phys. Rev. B}\ }\textbf {\bibinfo {volume}
  {66}},\ \bibinfo {pages} {205104} (\bibinfo {year} {2002})}\BibitemShut
  {NoStop}%
\bibitem [{\citenamefont {Kogut}(1979)}]{Kogut1979}%
  \BibitemOpen
  \bibfield  {author} {\bibinfo {author} {\bibfnamefont {J.~B.}\ \bibnamefont
  {Kogut}},\ }\href {\doibase 10.1103/RevModPhys.51.659} {\bibfield  {journal}
  {\bibinfo  {journal} {Rev. Mod. Phys.}\ }\textbf {\bibinfo {volume} {51}},\
  \bibinfo {pages} {659} (\bibinfo {year} {1979})}\BibitemShut {NoStop}%
\bibitem [{\citenamefont {Fradkin}(2013)}]{Fradkin2013}%
  \BibitemOpen
  \bibfield  {author} {\bibinfo {author} {\bibfnamefont {E.}~\bibnamefont
  {Fradkin}},\ }\href@noop {} {\emph {\bibinfo {title} {Field Theories of
  Condensed Matter Physics}}}\ (\bibinfo  {publisher} {Cambridge University
  Press, 2nd edition},\ \bibinfo {year} {2013})\BibitemShut {NoStop}%
\bibitem [{\citenamefont {Xu}\ \emph {et~al.}(2019{\natexlab{a}})\citenamefont
  {Xu}, \citenamefont {Liu}, \citenamefont {Pan}, \citenamefont {Qi},
  \citenamefont {Sun},\ and\ \citenamefont {Meng}}]{XiaoYanXu2019}%
  \BibitemOpen
  \bibfield  {author} {\bibinfo {author} {\bibfnamefont {X.~Y.}\ \bibnamefont
  {Xu}}, \bibinfo {author} {\bibfnamefont {Z.~H.}\ \bibnamefont {Liu}},
  \bibinfo {author} {\bibfnamefont {G.}~\bibnamefont {Pan}}, \bibinfo {author}
  {\bibfnamefont {Y.}~\bibnamefont {Qi}}, \bibinfo {author} {\bibfnamefont
  {K.}~\bibnamefont {Sun}}, \ and\ \bibinfo {author} {\bibfnamefont {Z.~Y.}\
  \bibnamefont {Meng}},\ }\href {\doibase 10.1088/1361-648x/ab3295} {\bibfield
  {journal} {\bibinfo  {journal} {Journal of Physics: Condensed Matter}\
  }\textbf {\bibinfo {volume} {31}},\ \bibinfo {pages} {463001} (\bibinfo
  {year} {2019}{\natexlab{a}})}\BibitemShut {NoStop}%
\bibitem [{\citenamefont {Xu}\ \emph {et~al.}(2019{\natexlab{b}})\citenamefont
  {Xu}, \citenamefont {Qi}, \citenamefont {Zhang}, \citenamefont {Assaad},
  \citenamefont {Xu},\ and\ \citenamefont {Meng}}]{XiaoYanXu2018}%
  \BibitemOpen
  \bibfield  {author} {\bibinfo {author} {\bibfnamefont {X.~Y.}\ \bibnamefont
  {Xu}}, \bibinfo {author} {\bibfnamefont {Y.}~\bibnamefont {Qi}}, \bibinfo
  {author} {\bibfnamefont {L.}~\bibnamefont {Zhang}}, \bibinfo {author}
  {\bibfnamefont {F.~F.}\ \bibnamefont {Assaad}}, \bibinfo {author}
  {\bibfnamefont {C.}~\bibnamefont {Xu}}, \ and\ \bibinfo {author}
  {\bibfnamefont {Z.~Y.}\ \bibnamefont {Meng}},\ }\href {\doibase
  10.1103/PhysRevX.9.021022} {\bibfield  {journal} {\bibinfo  {journal} {Phys.
  Rev. X}\ }\textbf {\bibinfo {volume} {9}},\ \bibinfo {pages} {021022}
  (\bibinfo {year} {2019}{\natexlab{b}})}\BibitemShut {NoStop}%
\bibitem [{\citenamefont {Liu}\ \emph {et~al.}(2020)\citenamefont {Liu},
  \citenamefont {Wang}, \citenamefont {Sun},\ and\ \citenamefont
  {Meng}}]{YZLiu2020}%
  \BibitemOpen
  \bibfield  {author} {\bibinfo {author} {\bibfnamefont {Y.}~\bibnamefont
  {Liu}}, \bibinfo {author} {\bibfnamefont {W.}~\bibnamefont {Wang}}, \bibinfo
  {author} {\bibfnamefont {K.}~\bibnamefont {Sun}}, \ and\ \bibinfo {author}
  {\bibfnamefont {Z.~Y.}\ \bibnamefont {Meng}},\ }\href {\doibase
  10.1103/PhysRevB.101.064308} {\bibfield  {journal} {\bibinfo  {journal}
  {Phys. Rev. B}\ }\textbf {\bibinfo {volume} {101}},\ \bibinfo {pages}
  {064308} (\bibinfo {year} {2020})}\BibitemShut {NoStop}%
\bibitem [{\citenamefont {Hohenadler}\ and\ \citenamefont
  {Assaad}(2018)}]{Hohenadler2018}%
  \BibitemOpen
  \bibfield  {author} {\bibinfo {author} {\bibfnamefont {M.}~\bibnamefont
  {Hohenadler}}\ and\ \bibinfo {author} {\bibfnamefont {F.~F.}\ \bibnamefont
  {Assaad}},\ }\href {\doibase 10.1103/PhysRevLett.121.086601} {\bibfield
  {journal} {\bibinfo  {journal} {Phys. Rev. Lett.}\ }\textbf {\bibinfo
  {volume} {121}},\ \bibinfo {pages} {086601} (\bibinfo {year}
  {2018})}\BibitemShut {NoStop}%
\bibitem [{\citenamefont {Hohenadler}\ and\ \citenamefont
  {Assaad}(2019)}]{Hohenadler2019}%
  \BibitemOpen
  \bibfield  {author} {\bibinfo {author} {\bibfnamefont {M.}~\bibnamefont
  {Hohenadler}}\ and\ \bibinfo {author} {\bibfnamefont {F.~F.}\ \bibnamefont
  {Assaad}},\ }\href {\doibase 10.1103/PhysRevB.100.125133} {\bibfield
  {journal} {\bibinfo  {journal} {Phys. Rev. B}\ }\textbf {\bibinfo {volume}
  {100}},\ \bibinfo {pages} {125133} (\bibinfo {year} {2019})}\BibitemShut
  {NoStop}%
\bibitem [{\citenamefont {Xu}\ \emph {et~al.}(2017{\natexlab{a}})\citenamefont
  {Xu}, \citenamefont {Sun}, \citenamefont {Schattner}, \citenamefont {Berg},\
  and\ \citenamefont {Meng}}]{XiaoYanXu2017}%
  \BibitemOpen
  \bibfield  {author} {\bibinfo {author} {\bibfnamefont {X.~Y.}\ \bibnamefont
  {Xu}}, \bibinfo {author} {\bibfnamefont {K.}~\bibnamefont {Sun}}, \bibinfo
  {author} {\bibfnamefont {Y.}~\bibnamefont {Schattner}}, \bibinfo {author}
  {\bibfnamefont {E.}~\bibnamefont {Berg}}, \ and\ \bibinfo {author}
  {\bibfnamefont {Z.~Y.}\ \bibnamefont {Meng}},\ }\href {\doibase
  10.1103/PhysRevX.7.031058} {\bibfield  {journal} {\bibinfo  {journal} {Phys.
  Rev. X}\ }\textbf {\bibinfo {volume} {7}},\ \bibinfo {pages} {031058}
  (\bibinfo {year} {2017}{\natexlab{a}})}\BibitemShut {NoStop}%
\bibitem [{\citenamefont {Liu}\ \emph {et~al.}(2018)\citenamefont {Liu},
  \citenamefont {Xu}, \citenamefont {Qi}, \citenamefont {Sun},\ and\
  \citenamefont {Meng}}]{ZHLiu2018}%
  \BibitemOpen
  \bibfield  {author} {\bibinfo {author} {\bibfnamefont {Z.~H.}\ \bibnamefont
  {Liu}}, \bibinfo {author} {\bibfnamefont {X.~Y.}\ \bibnamefont {Xu}},
  \bibinfo {author} {\bibfnamefont {Y.}~\bibnamefont {Qi}}, \bibinfo {author}
  {\bibfnamefont {K.}~\bibnamefont {Sun}}, \ and\ \bibinfo {author}
  {\bibfnamefont {Z.~Y.}\ \bibnamefont {Meng}},\ }\href {\doibase
  10.1103/PhysRevB.98.045116} {\bibfield  {journal} {\bibinfo  {journal} {Phys.
  Rev. B}\ }\textbf {\bibinfo {volume} {98}},\ \bibinfo {pages} {045116}
  (\bibinfo {year} {2018})}\BibitemShut {NoStop}%
\bibitem [{\citenamefont {Liu}\ \emph {et~al.}(2019)\citenamefont {Liu},
  \citenamefont {Pan}, \citenamefont {Xu}, \citenamefont {Sun},\ and\
  \citenamefont {Meng}}]{ZHLiu2019}%
  \BibitemOpen
  \bibfield  {author} {\bibinfo {author} {\bibfnamefont {Z.~H.}\ \bibnamefont
  {Liu}}, \bibinfo {author} {\bibfnamefont {G.}~\bibnamefont {Pan}}, \bibinfo
  {author} {\bibfnamefont {X.~Y.}\ \bibnamefont {Xu}}, \bibinfo {author}
  {\bibfnamefont {K.}~\bibnamefont {Sun}}, \ and\ \bibinfo {author}
  {\bibfnamefont {Z.~Y.}\ \bibnamefont {Meng}},\ }\href {\doibase
  10.1073/pnas.1901751116} {\bibfield  {journal} {\bibinfo  {journal}
  {Proceedings of the National Academy of Sciences}\ } (\bibinfo {year}
  {2019}),\ 10.1073/pnas.1901751116}\BibitemShut {NoStop}%
\bibitem [{\citenamefont {Hirsch}(1985)}]{Hirsch1985}%
  \BibitemOpen
  \bibfield  {author} {\bibinfo {author} {\bibfnamefont {J.~E.}\ \bibnamefont
  {Hirsch}},\ }\href {\doibase 10.1103/PhysRevB.31.4403} {\bibfield  {journal}
  {\bibinfo  {journal} {Phys. Rev. B}\ }\textbf {\bibinfo {volume} {31}},\
  \bibinfo {pages} {4403} (\bibinfo {year} {1985})}\BibitemShut {NoStop}%
\bibitem [{\citenamefont {He}\ \emph {et~al.}(2016)\citenamefont {He},
  \citenamefont {Wu}, \citenamefont {You}, \citenamefont {Xu}, \citenamefont
  {Meng},\ and\ \citenamefont {Lu}}]{YYHe2016}%
  \BibitemOpen
  \bibfield  {author} {\bibinfo {author} {\bibfnamefont {Y.-Y.}\ \bibnamefont
  {He}}, \bibinfo {author} {\bibfnamefont {H.-Q.}\ \bibnamefont {Wu}}, \bibinfo
  {author} {\bibfnamefont {Y.-Z.}\ \bibnamefont {You}}, \bibinfo {author}
  {\bibfnamefont {C.}~\bibnamefont {Xu}}, \bibinfo {author} {\bibfnamefont
  {Z.~Y.}\ \bibnamefont {Meng}}, \ and\ \bibinfo {author} {\bibfnamefont
  {Z.-Y.}\ \bibnamefont {Lu}},\ }\href {\doibase 10.1103/PhysRevB.93.115150}
  {\bibfield  {journal} {\bibinfo  {journal} {Phys. Rev. B}\ }\textbf {\bibinfo
  {volume} {93}},\ \bibinfo {pages} {115150} (\bibinfo {year}
  {2016})}\BibitemShut {NoStop}%
\bibitem [{\citenamefont {Xu}\ \emph {et~al.}(2017{\natexlab{b}})\citenamefont
  {Xu}, \citenamefont {Beach}, \citenamefont {Sun}, \citenamefont {Assaad},\
  and\ \citenamefont {Meng}}]{XYXuSO42017}%
  \BibitemOpen
  \bibfield  {author} {\bibinfo {author} {\bibfnamefont {X.~Y.}\ \bibnamefont
  {Xu}}, \bibinfo {author} {\bibfnamefont {K.~S.~D.}\ \bibnamefont {Beach}},
  \bibinfo {author} {\bibfnamefont {K.}~\bibnamefont {Sun}}, \bibinfo {author}
  {\bibfnamefont {F.~F.}\ \bibnamefont {Assaad}}, \ and\ \bibinfo {author}
  {\bibfnamefont {Z.~Y.}\ \bibnamefont {Meng}},\ }\href {\doibase
  10.1103/PhysRevB.95.085110} {\bibfield  {journal} {\bibinfo  {journal} {Phys.
  Rev. B}\ }\textbf {\bibinfo {volume} {95}},\ \bibinfo {pages} {085110}
  (\bibinfo {year} {2017}{\natexlab{b}})}\BibitemShut {NoStop}%
\bibitem [{\citenamefont {Assaad}\ and\ \citenamefont
  {Grover}(2016)}]{Assaad2016}%
  \BibitemOpen
  \bibfield  {author} {\bibinfo {author} {\bibfnamefont {F.~F.}\ \bibnamefont
  {Assaad}}\ and\ \bibinfo {author} {\bibfnamefont {T.}~\bibnamefont
  {Grover}},\ }\href {\doibase 10.1103/PhysRevX.6.041049} {\bibfield  {journal}
  {\bibinfo  {journal} {Phys. Rev. X}\ }\textbf {\bibinfo {volume} {6}},\
  \bibinfo {pages} {041049} (\bibinfo {year} {2016})}\BibitemShut {NoStop}%
\bibitem [{\citenamefont {Gazit}\ \emph {et~al.}(2018)\citenamefont {Gazit},
  \citenamefont {Assaad}, \citenamefont {Sachdev}, \citenamefont {Vishwanath},\
  and\ \citenamefont {Wang}}]{Gazit2018}%
  \BibitemOpen
  \bibfield  {author} {\bibinfo {author} {\bibfnamefont {S.}~\bibnamefont
  {Gazit}}, \bibinfo {author} {\bibfnamefont {F.~F.}\ \bibnamefont {Assaad}},
  \bibinfo {author} {\bibfnamefont {S.}~\bibnamefont {Sachdev}}, \bibinfo
  {author} {\bibfnamefont {A.}~\bibnamefont {Vishwanath}}, \ and\ \bibinfo
  {author} {\bibfnamefont {C.}~\bibnamefont {Wang}},\ }\href {\doibase
  10.1073/pnas.1806338115} {\bibfield  {journal} {\bibinfo  {journal}
  {Proceedings of the National Academy of Sciences}\ }\textbf {\bibinfo
  {volume} {115}},\ \bibinfo {pages} {E6987} (\bibinfo {year}
  {2018})}\BibitemShut {NoStop}%
\bibitem [{\citenamefont {Gazit}\ \emph {et~al.}(2017)\citenamefont {Gazit},
  \citenamefont {Randeria},\ and\ \citenamefont {Vishwanath}}]{Gazit2016}%
  \BibitemOpen
  \bibfield  {author} {\bibinfo {author} {\bibfnamefont {S.}~\bibnamefont
  {Gazit}}, \bibinfo {author} {\bibfnamefont {M.}~\bibnamefont {Randeria}}, \
  and\ \bibinfo {author} {\bibfnamefont {A.}~\bibnamefont {Vishwanath}},\
  }\href {http://dx.doi.org/10.1038/nphys4028} {\bibfield  {journal} {\bibinfo
  {journal} {Nat Phys}\ }\textbf {\bibinfo {volume} {advance online
  publication}} (\bibinfo {year} {2017})}\BibitemShut {NoStop}%
\end{thebibliography}%

\newpage




\appendix
\section{Quantum Monte Carlo Implementation}
After descritizing the imaginary time $\beta=\Delta\tau L_{\tau}$,
and performing the trace of Ising matter degrees of freedom in the
$S^{z}$ basis, tracing of $\mathbb{Z}_2$ gauge field degrees of freedom in
$\sigma^{z}$ basis, and tracing of fermion degrees of freedom in
the occupation number basis, the partition function can be written
as
\begin{align}
Z= & \text{Tr}\left(e^{-\beta H}\right)\\
= & \exp\left[\sum_{l,\langle i,j\rangle}\Delta\tau JS_{i}^{z}(l)\sigma_{b}^{z}(l)S_{j}^{z}(l)+\sum_{i,\langle l,l'\rangle}\gamma_{s}S_{i}^{z}(l)S_{i}^{z}(l')\right]\times \nonumber \\
& \exp\left[\sum_{l,\square}\Delta\tau K\prod_{b\in\square}\sigma_{b}^{z}(l)+\sum_{b,\langle l,l'\rangle}\gamma_{\sigma}\sigma_{b}^{z}(l)\sigma_{b}^{z}(l')\right]\times \nonumber \\
& \left[\det\left(I+\prod_{l=L_{\tau}}^{1}\exp\left(V(l)\right)\right)\right]^{2},
\end{align}
where $\gamma_{s}=-\frac{1}{2}\ln\left(\tanh(\Delta\tau h)\right)$,
$\gamma_{\sigma}=-\frac{1}{2}\ln\left(\tanh(\Delta\tau g)\right)$,
and matrices $V(l)$ (imaginary time-slice index $l$ takes values $1,\cdots, L_\tau$) have elements $V(l)_{\langle i,j\rangle}=\Delta\tau t\sigma_{b}^{z}(l)$
and $V(l)_{i,i}=\Delta\tau\mu$. The square outside of the determinant
comes from two speicies of fermion (spin up and down). As the bosonic
part of weights are always positive, and the fermion part of weight
is a square of determinant of real matrix, the whole weight will be
always semi-positive, and it is absence of sign problem.

We will use determinant quantum Monte Carlo to simulate this model,
which has been widely used in simulating fermion boson coupled lattice
models and more details can be find in Refs.~\cite{XiaoYanXu2019}. The local updates
are performed on the Ising matter field $\{S_{i}^{z}\}$ and $\mathbb{Z}_{2}$
gauge fields $\{\sigma_{b}^{z}\}$ in a space-time configurational
space with volume $L_{\tau}\times L\times L$, where $L_{\tau}=\beta/\Delta\tau$
with $\Delta\tau=0.1$ and $\beta=L=$12, 14, ..., 20, 24.

\section{Away from Half-Filling}
\begin{figure*}[htp!]
	\begin{center}
		\includegraphics[width=0.8\textwidth]{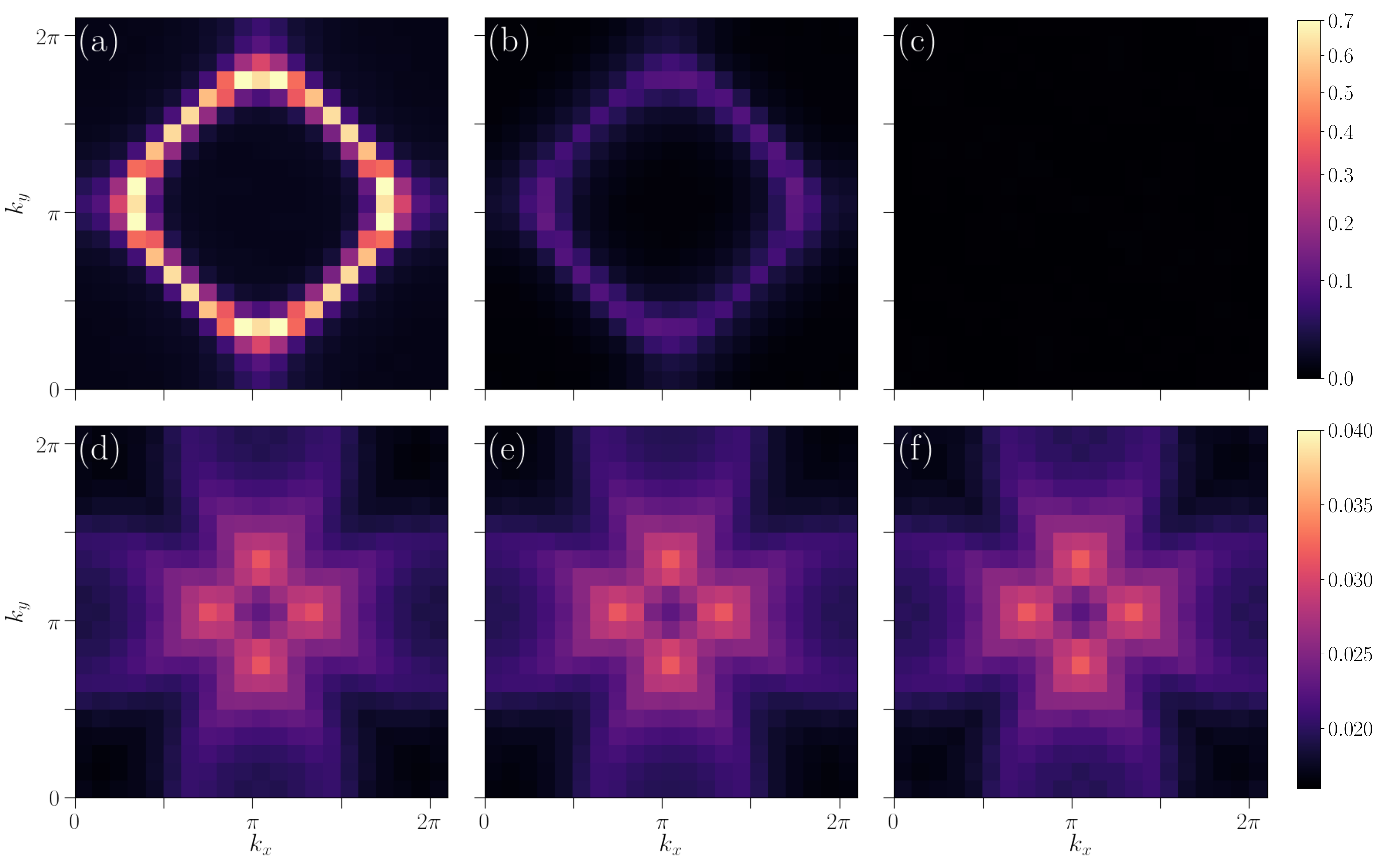}
		\caption{Fermionic and bosonic responses at filling 0.7. (a), (b) and (c): the spectra at the FS $A(\mathbf{k},\omega=0)\propto G(\mathbf{k},\beta/2)$ of composite fermion $c$ for $L=20$, $T=0.1$, $g=0.5$ systems. For $h=0.5$ ((a), $h<h_c$, NM), $h=2$ ((b), $h\sim h_c$, QCP) and $h=4$ ((c), $h>h_c$, OM). (d), (e) and (f): spin susceptibility $\chi(\mathbf{q},\omega=0)$ for the same parameter sets. It is clear that in the NM phase (a), the near diamond shape FS gives rise to the magnetic instability a bit away from $\mathbf{q}=(\pi,\pi)$ in (d), but as the NM evolves into OM, the FS vanishes as shown in  (b) and (c), its magnetic response does not change in any obvious way, (e) and (f), that, there still exists the instability at same positions of (d) despite of the fact that there is no FS any longer at QCP and in OM phase. This is the special properties of the OM that it responds like a metal (magnetically and electronically) but there is a gap in its $A(\mathbf{k},\omega=0)$.}
		\label{fig:figS1}
	\end{center}
\end{figure*}

In this section, we provide the evolution of the FS away from the half-filled case. In Fig.~\ref{fig:figS1} the filling of the $f$-fermion is at $0.7$ with $L=20$ and $T=0.1$, and all the other parameters are the same with those in Fig.~\ref{fig:fig2} of the main text. Fig.~\ref{fig:figS1} (a) and (d) are the $A(\mathbf{k},\omega=0)$ and $\chi(\mathbf{q},\omega=0)$ at $h=0.5$. The system is inside the NM phase and since the nesting condition of the FS is less ideal, the magnetic susceptibility has maximal shifted slightly away from the $\mathbf{q}=(\pi,\pi)$. When $h=2.0$, close to the NM-OM transition, the $A(\mathbf{k},\omega=0)$ becomes very small [Fig.~\ref{fig:figS1} (b)], signifying the losing of coherence in the gauge-neutral $c$-fermions, but the its magnetic response still persists [Fig.~\ref{fig:figS1} (e)]. And when the $h=4.0$, the system is well inside the OM phase, Fig.~\ref{fig:figS1} (c) demonstrate the vanishing of the FS without any form of symmetry-breaking, i.e., violation of the Luttinger's theorem, and Fig.~\ref{fig:figS1} (f) depicts the same magnetic response as those of Fig.~\ref{fig:figS1} (d) and (e), namely, although the OM phase is an insulator from the single-particle perspective, it still reacts toward external perturbation as if it were metal.

In any way, the NM-OM transition shown in Fig.~\ref{fig:figS1} is more robust than that of Fig.~\ref{fig:fig2} in the main text, as the magnetic instability is actually weaker at filling 0.7 than the perfectly nested case at half-filling, which supports even strongly to the existence of the OM phase and the non-trivial QCP between it and the NM phase.

\end{document}